\documentclass[]{pasj01}

\begin{document} 
\Received{}
\Accepted{}

\title{MAXI observations of long X-ray bursts}

\author{Motoko \textsc{Serino}\altaffilmark{1}}%
\altaffiltext{1}
{MAXI team, RIKEN, 2-1 Hirosawa, Wako, Saitama 351-0198, Japan}
\email{motoko@crab.riken.jp}

\author{Wataru \textsc{Iwakiri}\altaffilmark{1}}

\author{Toru \textsc{Tamagawa}\altaffilmark{2}}
\altaffiltext{2}
{Nishina Center, RIKEN, 2-1 Hirosawa, Wako, Saitama 351-0198, Japan}

\author{Takanori \textsc{Sakamoto}\altaffilmark{3}}
\altaffiltext{3}{Department of Physics and Mathematics,
Aoyama Gakuin University,\\ 5-10-1 Fuchinobe, Chuo-ku, 
Sagamihara, Kanagawa 252-5258}

\author{Satoshi \textsc{Nakahira}\altaffilmark{4}}
\altaffiltext{4}{JEM Mission Operations and Integration Center, 
Human Spaceflight Technology Directorate, Japan Aerospace Exploration 
Agency, 2-1-1 Sengen, Tsukuba, Ibaraki 305-8505, Japan}

\author{Masaru \textsc{Matsuoka}\altaffilmark{1}}

\author{Kazutaka \textsc{Yamaoka}\altaffilmark{5,6}}
\altaffiltext{5}
{Department of Particle Physics and Astronomy, Nagoya University, 
Furo-cho, Chikusa-ku, Nagoya, Aichi 464-8601, Japan}
\altaffiltext{6}
{Solar-Terrestrial Environment Laboratory, Nagoya University,
Furo-cho, Chikusa-ku, Nagoya, Aichi 464-8601, Japan}

\author{Hitoshi \textsc{Negoro}\altaffilmark{7}}
\altaffiltext{7}{Department of Physics, Nihon University,
1-8-14 Kanda-Surugadai, Chiyoda-ku, Tokyo 101-8308, Japan}



\maketitle

\begin{abstract}

We report nine long X-ray bursts from neutron stars, detected with 
Monitor of All-sky X-ray Image (MAXI).
Some of these bursts lasted for hours, and hence are qualified as
superbursts,
which are prolonged thermonuclear flashes on neutron stars
and are relatively rare events. 
MAXI observes roughly 85\% of the whole sky every 92 minutes 
in the 2--20 keV energy band, and has detected nine
bursts with a long e-folding decay time, ranging from 0.27 to 5.2 hours, 
since its launch in 2009 August until 2015 August.
The majority of the nine events were found to originate from transient
X-ray sources.
The persistent luminosities of the sources, when these prolonged bursts 
were observed, were lower than 1\% of the Eddington luminosity for 
five of them and lower than 20\% for the rest.
This trend is contrastive to the 18 superbursts observed before MAXI,
all but two of which originated from bright persistent sources.
The distribution of the total emitted energy, 
i.e., the product of e-folding time and luminosity,
of these bursts clusters around 10$^{41}$--10$^{42}$ erg,
whereas either of the e-folding time and luminosity ranges for an order of magnitude.
Among the nine events, two were from 4U 1850$-$086 during the phases
of relatively low persistent-flux, whereas it usually exhibits
standard short X-ray bursts during outbursts.
\end{abstract}

\section{Introduction}

Since the discovery of  a prolonged thermonuclear X-ray burst 
with the duration of several hours
from 4U 1735$-$44 (\cite{2000A&A...357L..21C}),
about 24 of such bursts, referred to as ``superbursts'' \citep{2001ApJ...554L..59W}, have been reported from 14 sources
\citep{2012ApJ...752..150K,2012ATel.4622....1N,2014ATel.6668....1S}
.
Among those, more than one superburst have been observed from 
4U 1636$-$536, GX 17+2, and Ser X-1 
\citep{2001ApJ...554L..59W,2004AIPC..714..257K,2009ATel.2140....1K}.
The shortest observed recurrence time of superbursts was only 8 days in 
GX 17+2
\citep{2004A&A...426..257I}.

Superbursts are considered to be thermonuclear flashes on neutron stars
in low-mass X-ray binaries,
 and are similar to  ``normal'' type-I X-ray bursts with a 
shorter duration. In fact, all the superburst sources are known
 to be (normal) X-ray burst sources.
The  primary difference  between normal and superbursts is the duration.
There is also a class of X-ray bursts with
``intermediate duration'' \citep{2006ApJ...646..429C},
which have the decay time ranging from a hundred to a thousand 
seconds
\citep{2008A&A...484...43F}.
It is thought that the duration of an X-ray burst is related to 
the depth of the ignition point and the composition of the fuel. 
Superbursts are ignited by carbon burning deep in an ocean of heavier 
elements on a neutron star \citep{2001ApJ...559L.127C}.
On the other hand, intermediate-duration bursts are ignited by a thick
layer of helium,
which is most commonly provided by a degenerate companion star 
\citep{2005A&A...441..675I,2008A&A...484...43F}.

The composition of the fuel depends on 
the mass accretion rate.
The relation between the accretion rate and the time
scale of the burst has been studied for decades
(e.g. \cite{1981ApJ...247..267F} for normal bursts).
Empirically, superbursts 
occurred on the persistent (not transient) sources with a
high  luminosity ($>$10\% of the Eddington luminosity)
\citep{2004NuPhS.132..466K,2004A&A...426..257I},
whereas intermediate-duration bursts can occur in the sources
with a very low luminosity ($<$1\% of the Eddington luminosity)
\citep{2005A&A...441..675I,2008A&A...484...43F}.

However, it is now understood that superbursts can occur in
transient sources and persistent sources with a low luminosity
\citep{2011ATel.3183....1C,2011ATel.3760....1A,2012PASJ...64...91S,
2014ATel.6668....1S}.
For example, a superburst was observed from a transient source, 
4U 1608$-$522 \citep{2008A&A...479..177K}.
The superburst occurred during an 
outburst, which  had started 57.6 days before the burst.
\authorcite{2008A&A...479..177K} argued that the fuel of the
superburst accumulated during the outburst was not sufficient
to generate the released energy in the burst.
4U 1608$-$522 had shown multiple outbursts before the superburst.
\authorcite{2008A&A...479..177K} concluded that the fuel
should have been accumulated during the preceding multiple outbursts.
Another example is 4U~0614+091, a source with a low persistent 
luminosity ($<$1\% of Eddington limit), 
from which a superburst has been detected on 2005 March 12 (MJD 53441)
\citep{2010A&A...514A..65K}.
Although intermediate-duration bursts from 4U~0614+091
had been detected,
superbursts from such a low-luminosity source had not been
observed before that event.
These discoveries presented serious problems in theoretical models of
superburst ignition \citep{2012MNRAS.426..927A}.

The Gas Slit Camera (GSC; \cite{2011PASJ...63S.623M}) onboard
Monitor of All-sky X-ray Image (MAXI; \cite{2009PASJ...61..999M})
observes about 85\% of the whole sky every 92 minutes
\citep{2011PASJ...63S.635S}, and has the capability of detecting
transient events with the detection limit of 
 $\sim$2 $\times 10^{-9}$ erg cm$^{-2}$ s$^{-1}$ in the 2--20 keV band
(e.g. \cite{2014PASJ...66...87S,negoro.ns}) in a scan transit.
This sensitivity is sufficient to detect emission of superbursts
from galactic neutron-star binaries.
The typical interval of 92 min between two GSC scans is short enough 
not to miss bright transient events which last for several hours,
such as superbursts, in almost the entire sky.
 Indeed, all the superbursts reported since 2009 till the time of this 
 writing have been observed by MAXI.

In this paper we present long-lasting X-ray bursts detected with MAXI, 
which were observed in more than a scan transit 
(typically 40--50 s, \cite{2011PASJ...63S.635S}).
We study nine long-lasting X-ray bursts and discuss their 
global properties. 
Eight out of nine bursts have
an e-folding decay time of the bolometric flux
much longer than a thousand seconds, which is supposed to be the upper
end of the intermediate-duration burst,
whereas the decay time of the other one was about a thousand seconds.
 We also look into the flux levels before and after the long bursts,
using the MAXI public light-curves of the sources.
In section \ref{ss:obs},
we describe the sample of long X-ray bursts observed  with MAXI and the
method of our analysis. The results  are  presented
in section \ref{ss:res}. In section \ref{ss:dis}, we discuss the
properties of individual events and their canonical characteristics,
and then summarize our results. 

\section{Observations}
\label{ss:obs}

\subsection{Sample in this paper}
MAXI has observed more than a hundred X-ray bursts%
\footnote{http://maxi.riken.jp/alert/novae/index.html}.
 Whereas some of them are found by human inspection,
most of the X-ray bursts are automatically detected by 
\textit{the MAXI Nova-Alert System}
\citep{negoro.ns}. The system works in real time and looks through 
the light curves of eight types of timescale bins (from 1 sec to 4 days)
at each celestial position. Among the timescale bins, 1 scan
(typically 40--50 sec) bin is useful to search for long X-ray bursts.
The system searches the newest time bin for a 3 sigma excess 
above the average flux over the previous 9 bins 
\citep{negoro.ns}.

If there is a known X-ray burst source within the error circle of a sky 
position, where an excess is found in the light curve, 
we regard it as an X-ray burst event.
In this paper we study the bursts whose the e-folding time is
longer than 100 seconds.
Seven events were detected in two or more consecutive scans, where the 
interval between scans is 92 min, 
and accordingly were selected as candidate long-lasting X-ray bursts.
Among them a superburst from SAX~J1747.0$-$2853 
(on 2011 February 13, or MJD 55605) was observed 
also by INTEGRAL \citep{2011ATel.3183....1C}. 

When a burst was detected in only one MAXI/GSC scan, which
lasts 40--50 s, it is unclear whether the e-folding time of the burst was
longer than 100 seconds or not, based on the MAXI/GSC data alone.
We in principle exclude those events from the sample in this paper
(for example, a burst from 4U~1850$-$086 on 2011 November 9 (MJD 55874)
in section \ref{ss:type}).

Two X-ray burst events are found to have a long duration, by combining 
the data of the MAXI/GSC and 
other satellites 
(See the column of ``inst.'' in table~\ref{tab:lst}).
One is a long X-ray burst from 4U 1850$-$086 on 2014 March 10 
(MJD 56726), which was observed by Swift \citep{2014ATel.5972....1I}.
For this source, the duration of the burst was too short to be 
observed in multiple scans of MAXI.
We did detect a significant increase of the fluxes in a MAXI scan 
during the burst duration expected from the Swift data.

The other one is a burst from 4U~1820$-$30 on 2010 March 17 (MJD 55272). 
MAXI failed to detect the excess of the flux in multiple scans 
but only in one.
 We nevertheless include it in our sample, 
 because the duration of the burst was at least 75 min,
 as reported by \citet{2011ATel.3625....1I}, whose estimation was based
 on both the flux of MAXI and the flux enhancement with the RXTE/ASM.

The burst observed on 2011 October 24 was from the direction toward
the globular cluster Terzan 5. Although MAXI cannot resolve the sources
in a globular cluster, we identified it as a burst from EXO~1745$-$248,
because the burst was followed by an outburst 
and the position of the outburst source was determined precisely
by Swift XRT and Chandra
\citep[and references there in]{2011ATel.3729....1M,2012PASJ...64...91S}. 
Table \ref{tab:lst} lists all the nine events.

\subsection{Data in this paper}
\begin{figure*}
 \begin{center}
   \includegraphics[width=8cm]{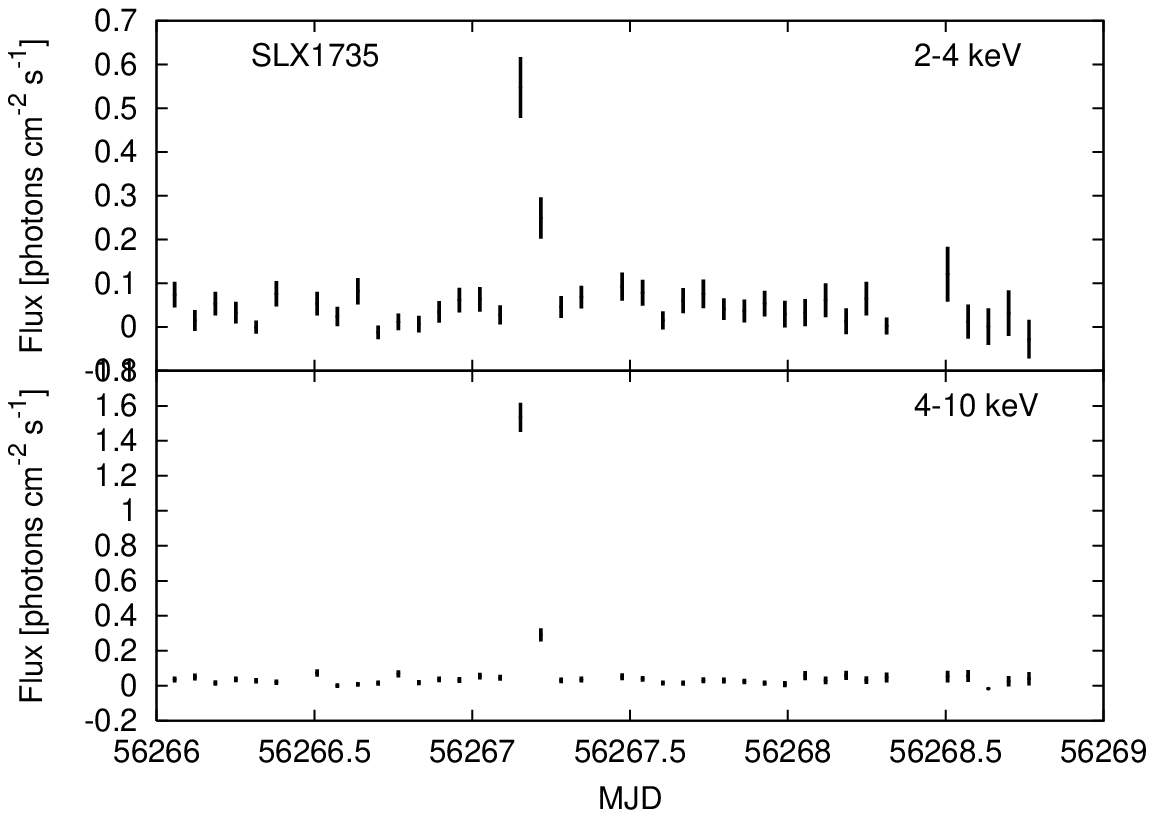}
   \includegraphics[width=8cm]{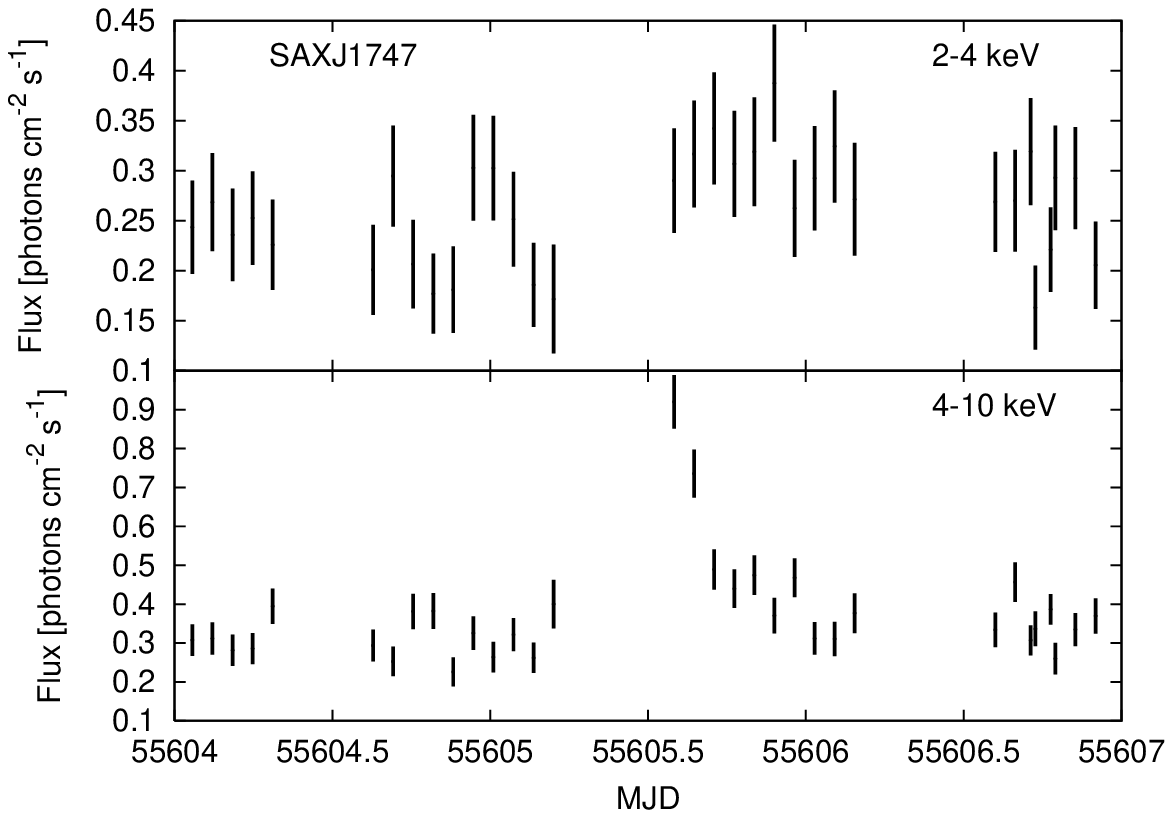}
   \includegraphics[width=8cm]{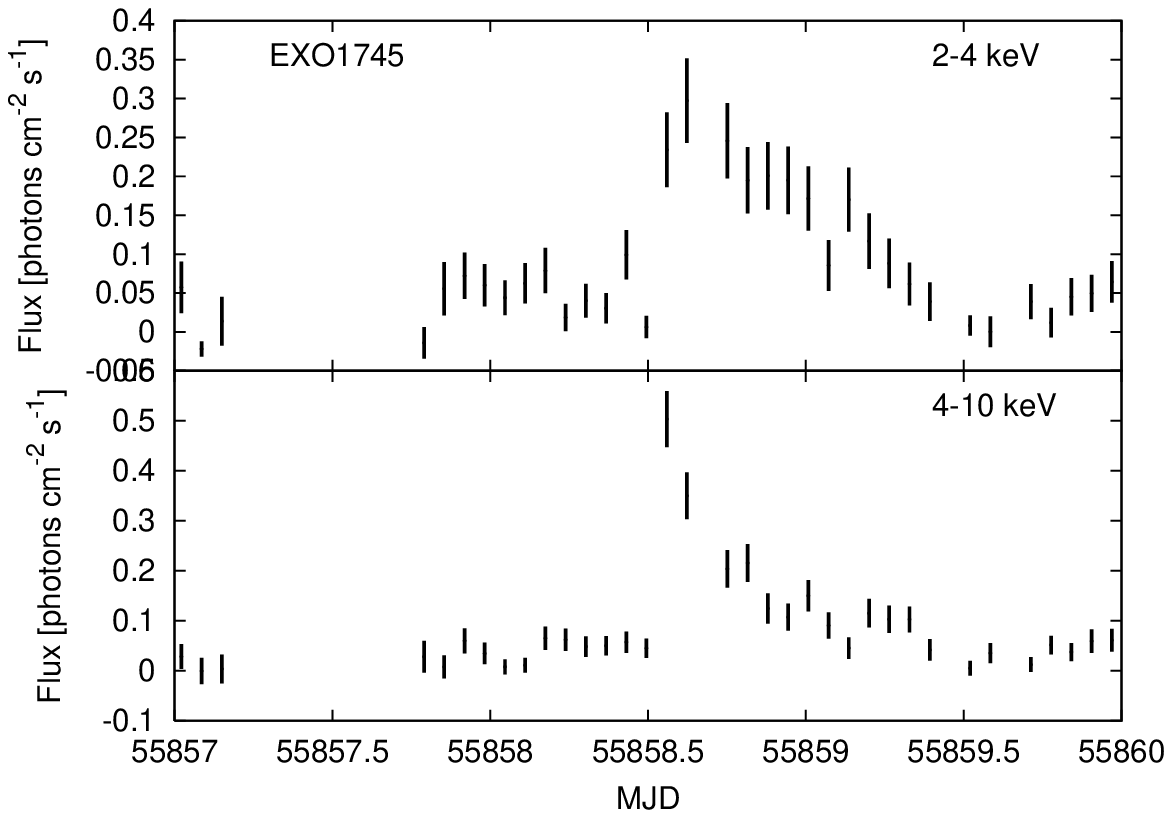}
   \includegraphics[width=8cm]{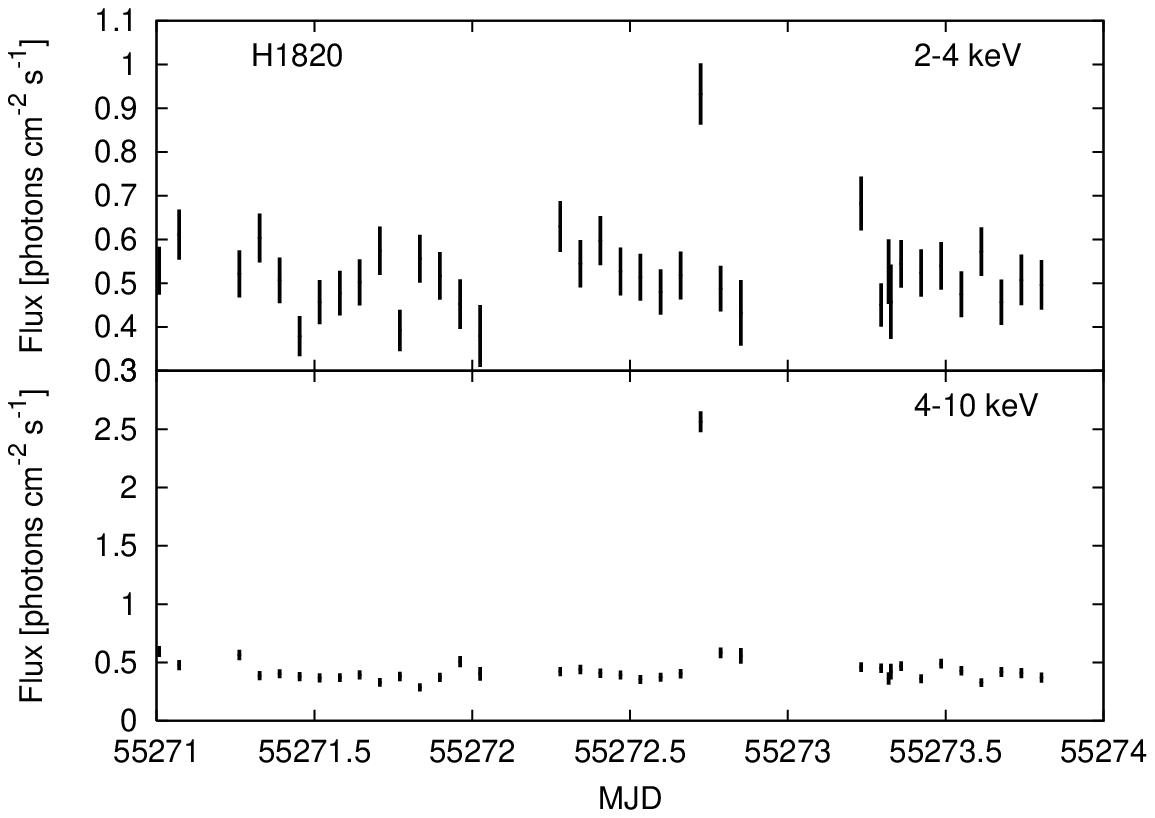}
   \includegraphics[width=8cm]{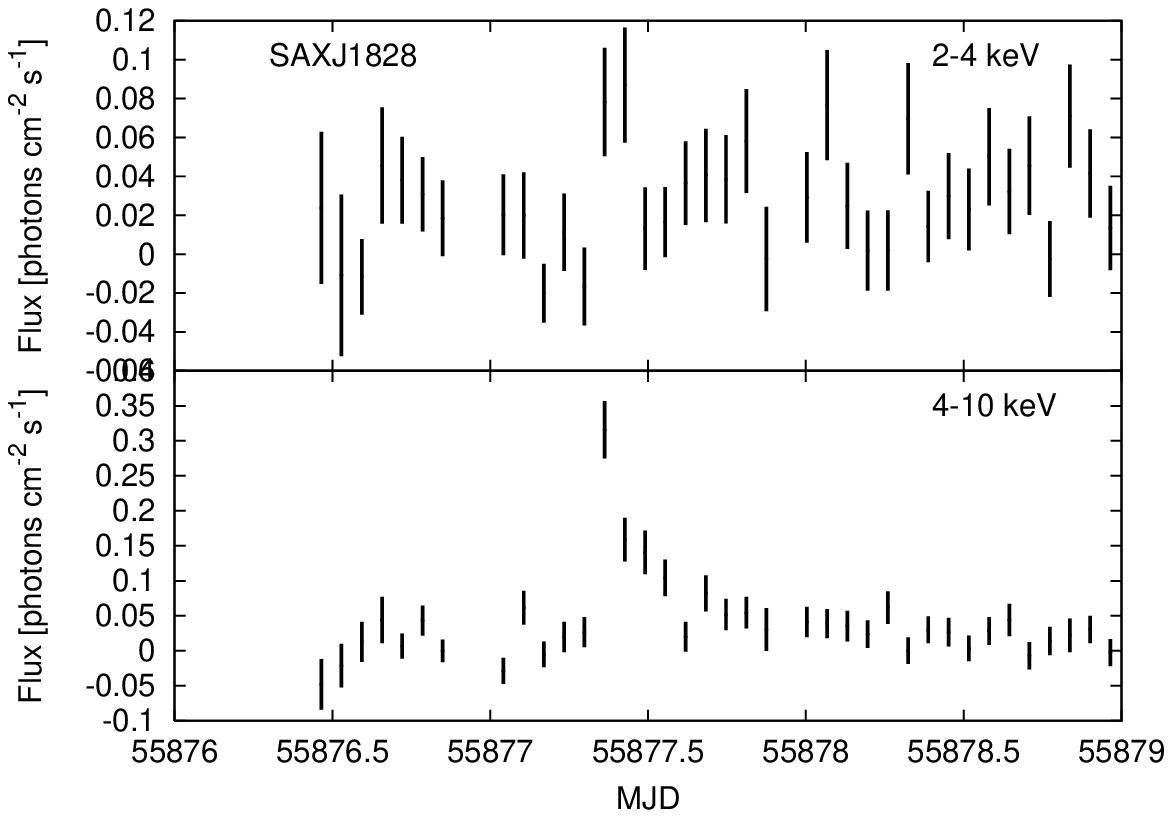}
   \includegraphics[width=8cm]{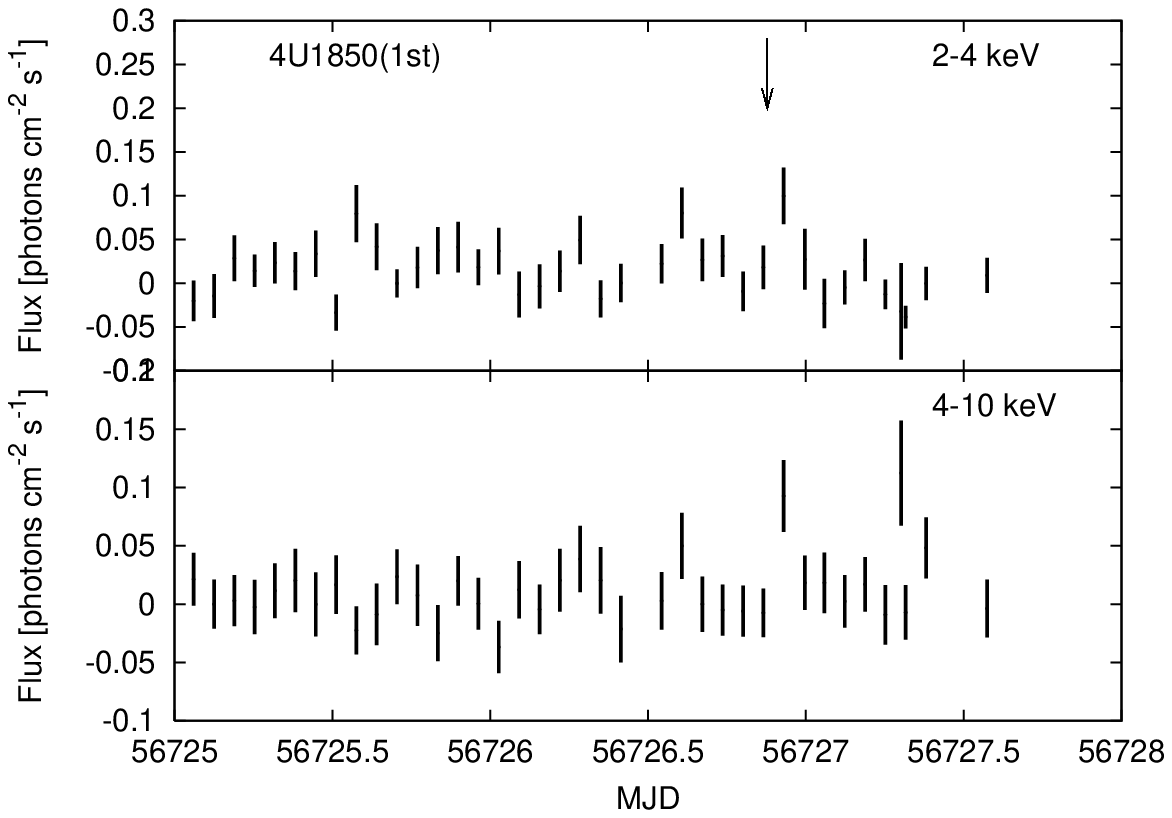}
   \includegraphics[width=8cm]{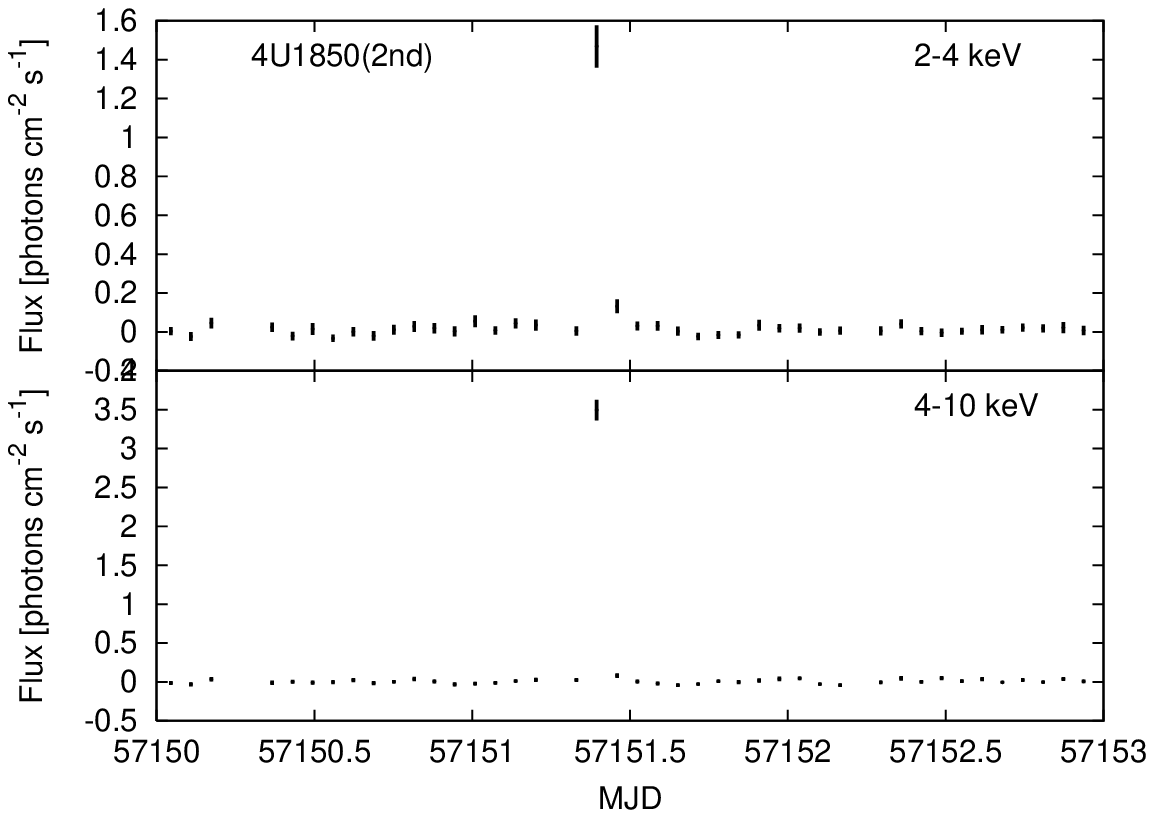}
   \includegraphics[width=8cm]{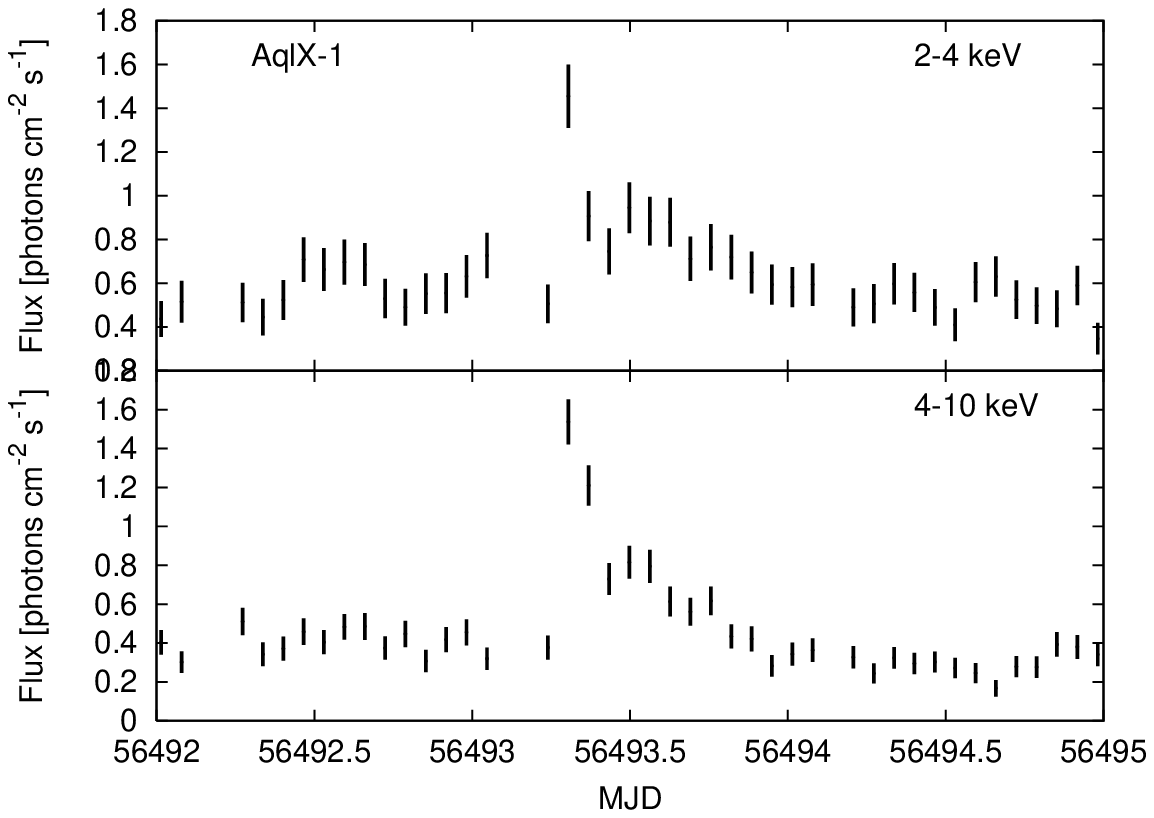}
 \end{center}
\caption{The light curves of the bursts in the 2--4 and 4--10 keV bands.
Each data point corresponds to a scan.
The typical time interval between two scans is 92 min.
For the first event from 4U 1850$-$086, the time of the Swift detection
is  indicated with  an arrow.
The light curve of 4U 0614+091 is not available around the burst time, 
because a structure of the International Space Station
 came close to the position of the source.
}
\label{fig:blc}
\end{figure*}

We use two types of data in the analyses.  
One is X-ray event data of MAXI GSC, which is used for the
spectral  and  time-series analyses within each scan.
The other is the MAXI public light-curves,%
  \footnote{http://maxi.riken.jp/top/}
 in which two separate time-resolutions (one orbit and
one day) are available for each object.
Figure \ref{fig:blc} shows the orbital light curves
of our sample in the 2--4 and 4--10 keV bands.%
\footnote{
The public light curve of 4U 0614+091 is not available
around the burst time, because  a structure of the  
International Space Station
 came close to the position of the source.}
 In general, the peaks of the bursts are more prominent in 4--10 keV
than in 2--4 keV bands and have a shorter decay time in the higher 
energy band.
For the study of persistent fluxes (section \ref{ss:persistent}),
we rebin the one-day light curves into ten-day bins in order to
study the persistent emissions of low flux level.
The results are shown in table~\ref{tab:pers}.

\subsection{Observations of normal bursts}
   In later sections we compare the fluxes of superbursts and normal 
   bursts.
   Table \ref{tab:pers} lists the peak bolometric flux of the normal 
   bursts in  the column $F_\mathrm{b}$.
  In the table, we mainly refer to the previous works  with various 
  satellites.
  The exception is 4U 1850$-$086, where we use the flux of a normal X-ray 
  burst observed  with MAXI on 2010 November 2 (MJD 55502)
  because the previous observation 
  with RXTE detected no X-ray burst from the source
  \citep{2008ApJS..179..360G}.
  The peak photon flux of  the burst  was 5.8$\pm$0.9 
  photons cm$^{-2}$ s$^{-1}$. Assuming a
  blackbody spectrum with the temperature of 2.0 keV,
  the peak bolometric  flux of the burst is calculated  to be
  5.9$\pm$0.9 $\times 10^{-8}$ erg cm$^{-2}$ s$^{-1}$.
  This peak flux with MAXI agreed
  with those of 28 normal bursts from this source in 2002,
  observed with HETE-2/WXM 
  \citep{2003AIPC..662....3R,2003SPIE.4851.1310S,2006NCimB.121.1593S}.
  Thus, the flux of the bursts has not changed much between 2002 and 2010.

\section{Data analysis and Results}
\label{ss:res}

\subsection{Variability within the scans of maximum fluxes}
\begin{figure*}
 \begin{center}
   \includegraphics[width=8cm]{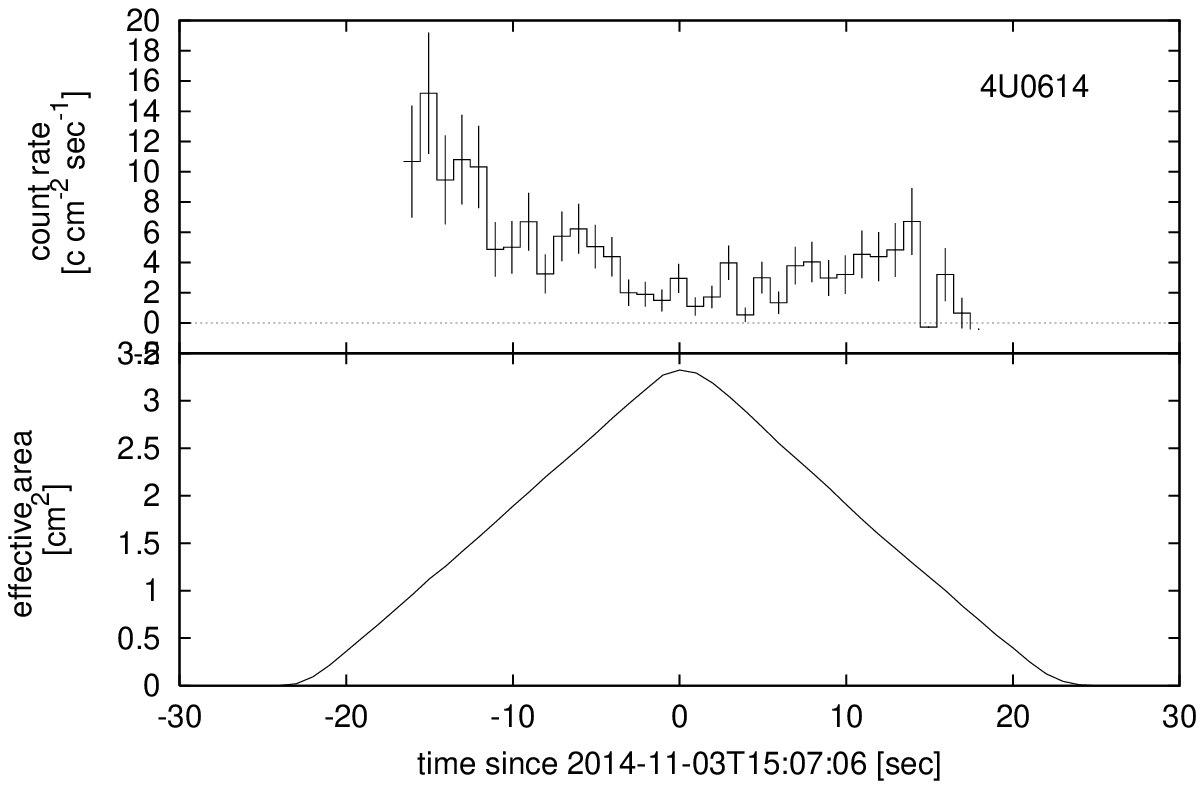}
   \includegraphics[width=8cm]{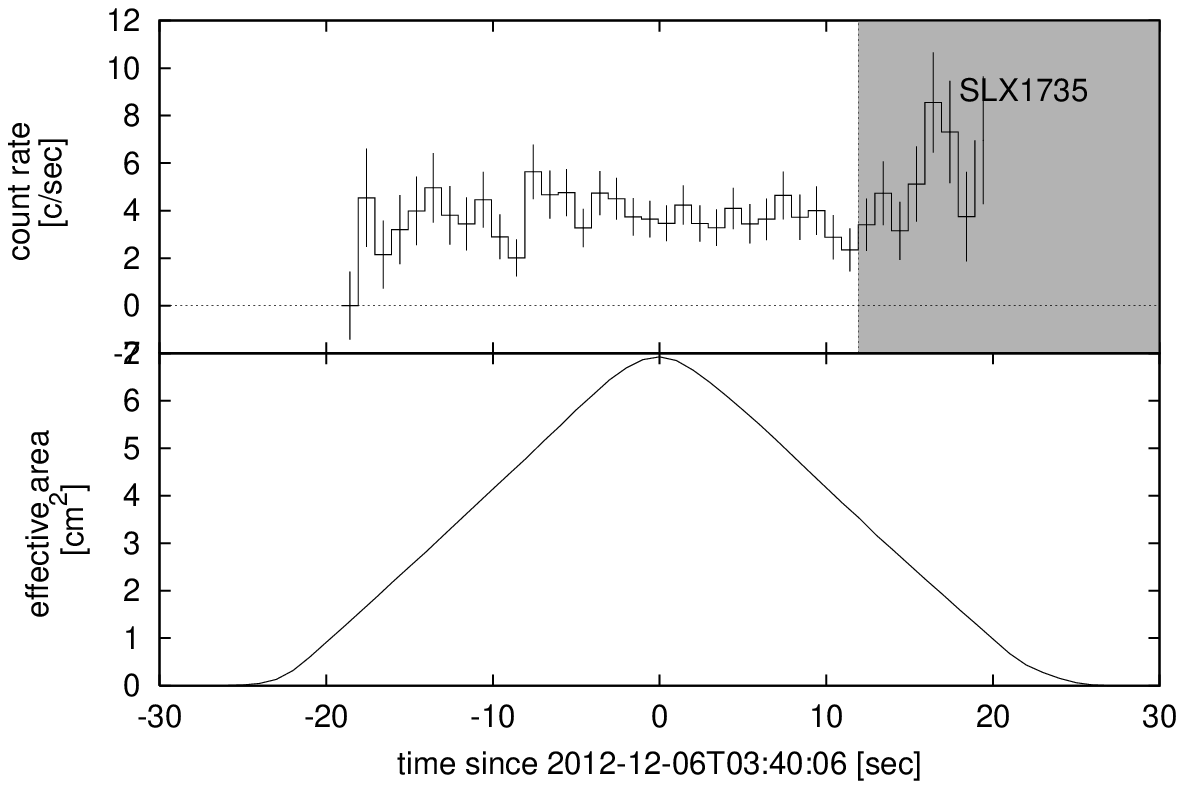}
   \includegraphics[width=8cm]{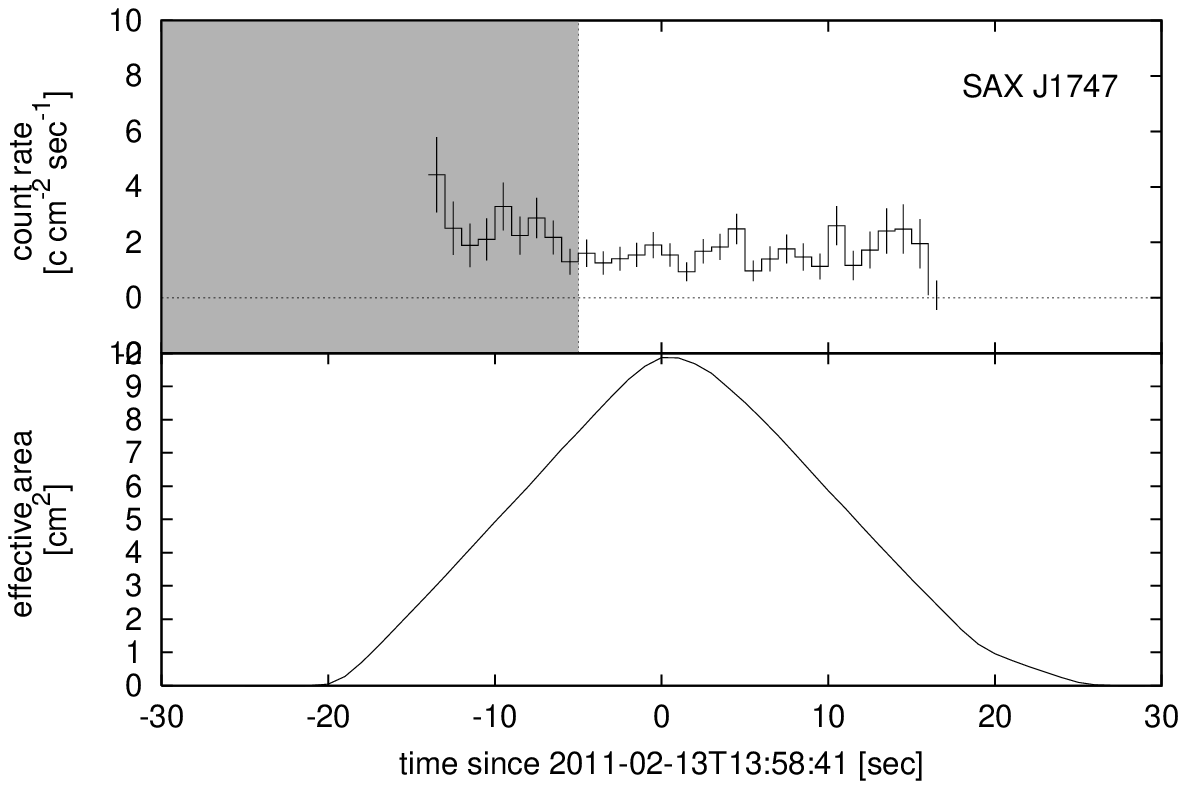}
   \includegraphics[width=8cm]{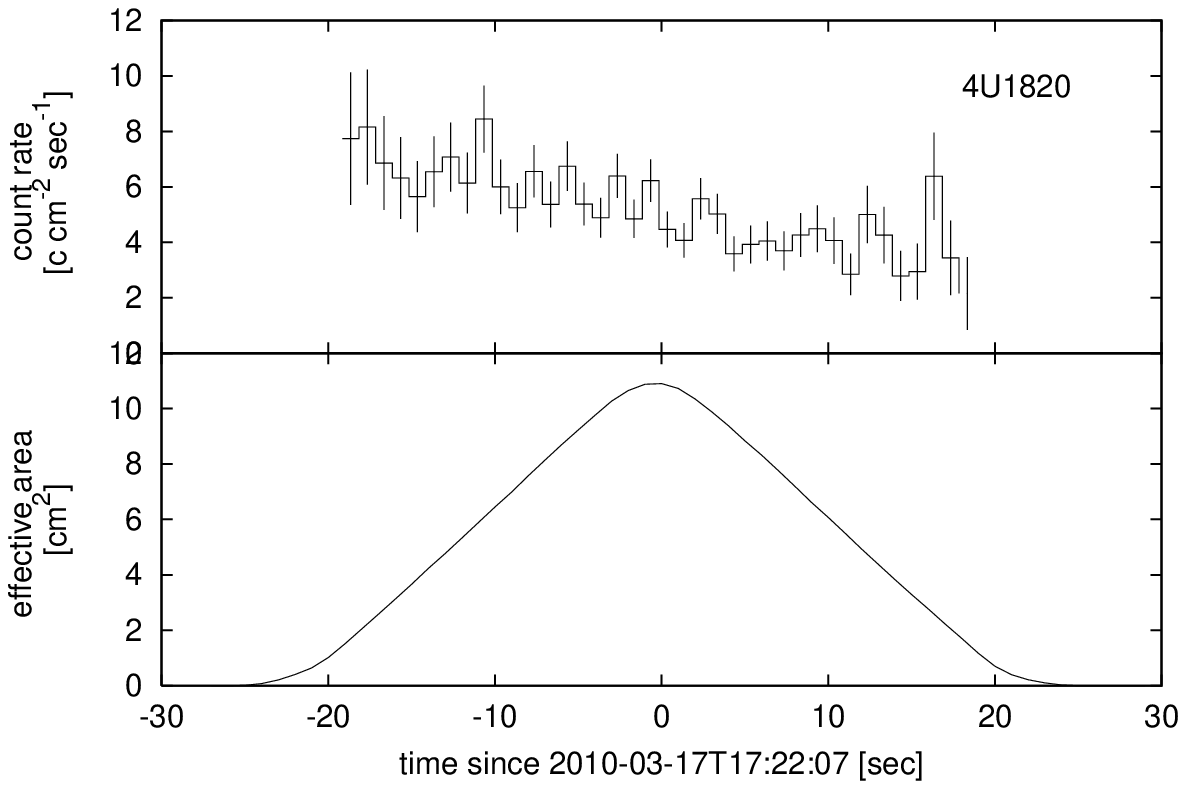}
   \includegraphics[width=8cm]{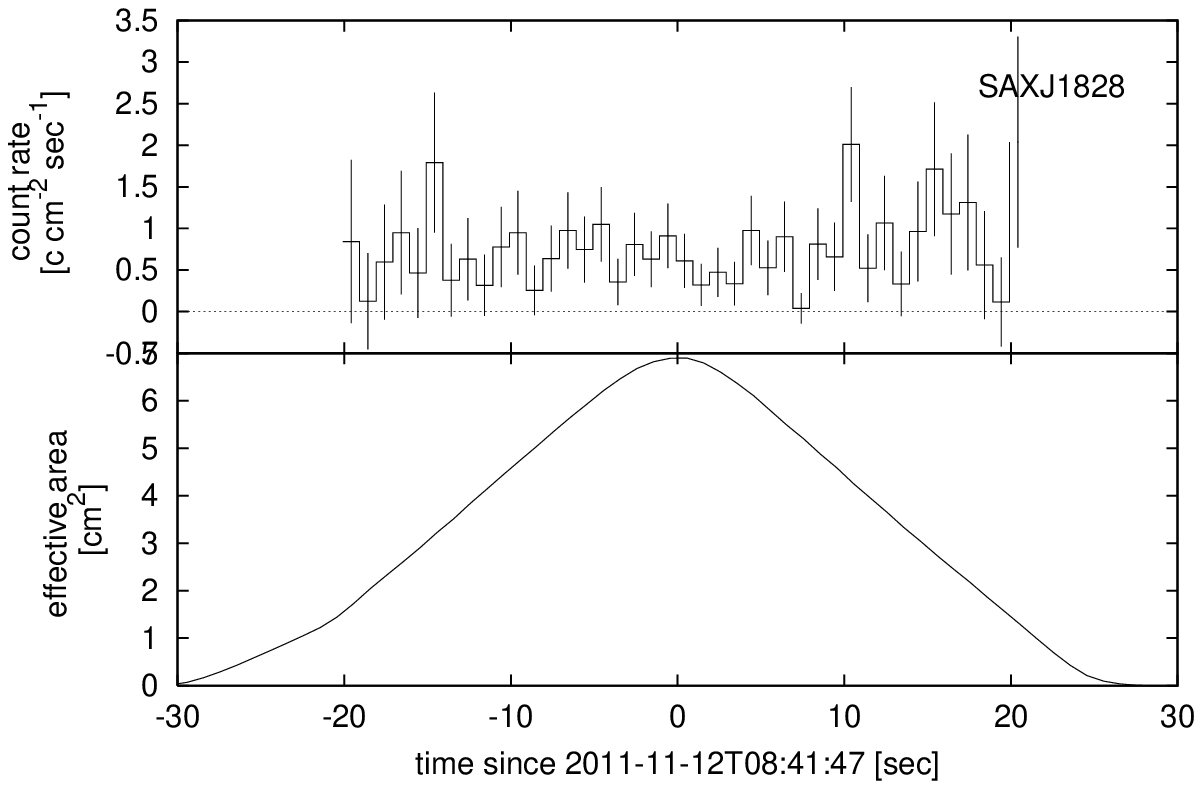}
   \includegraphics[width=8cm]{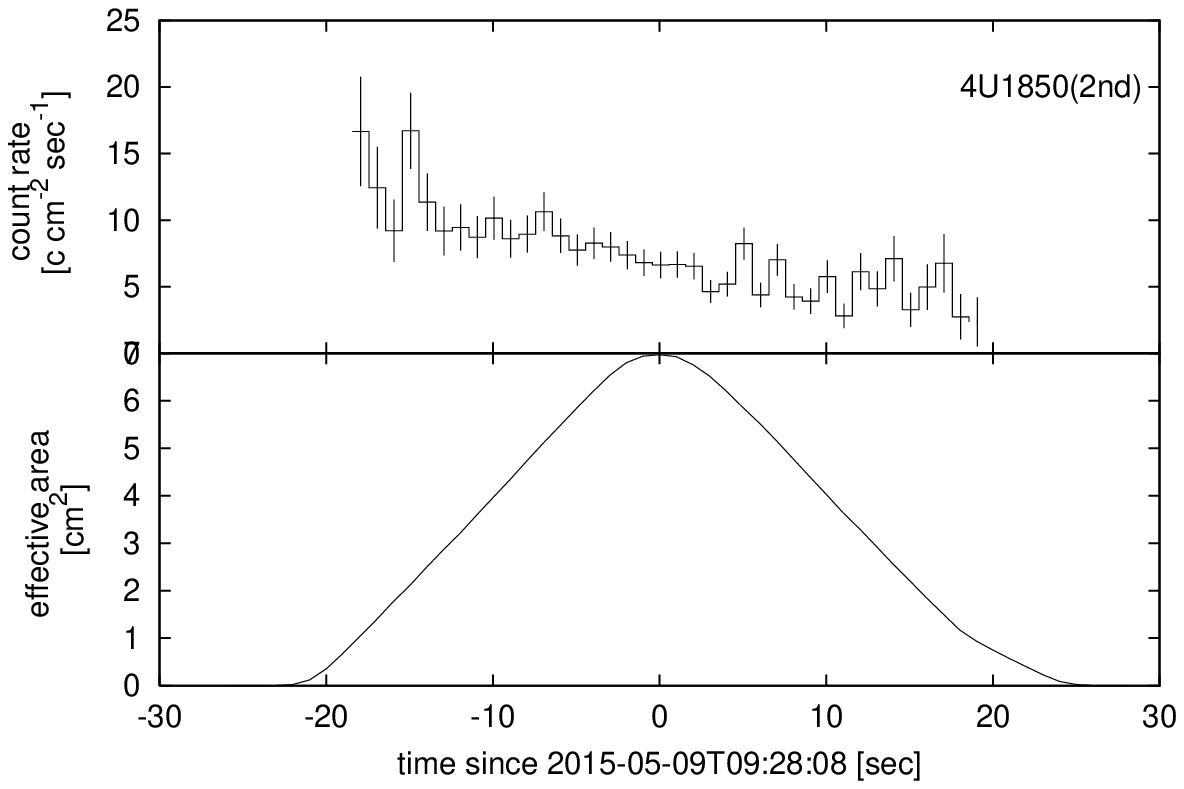}
   \includegraphics[width=8cm]{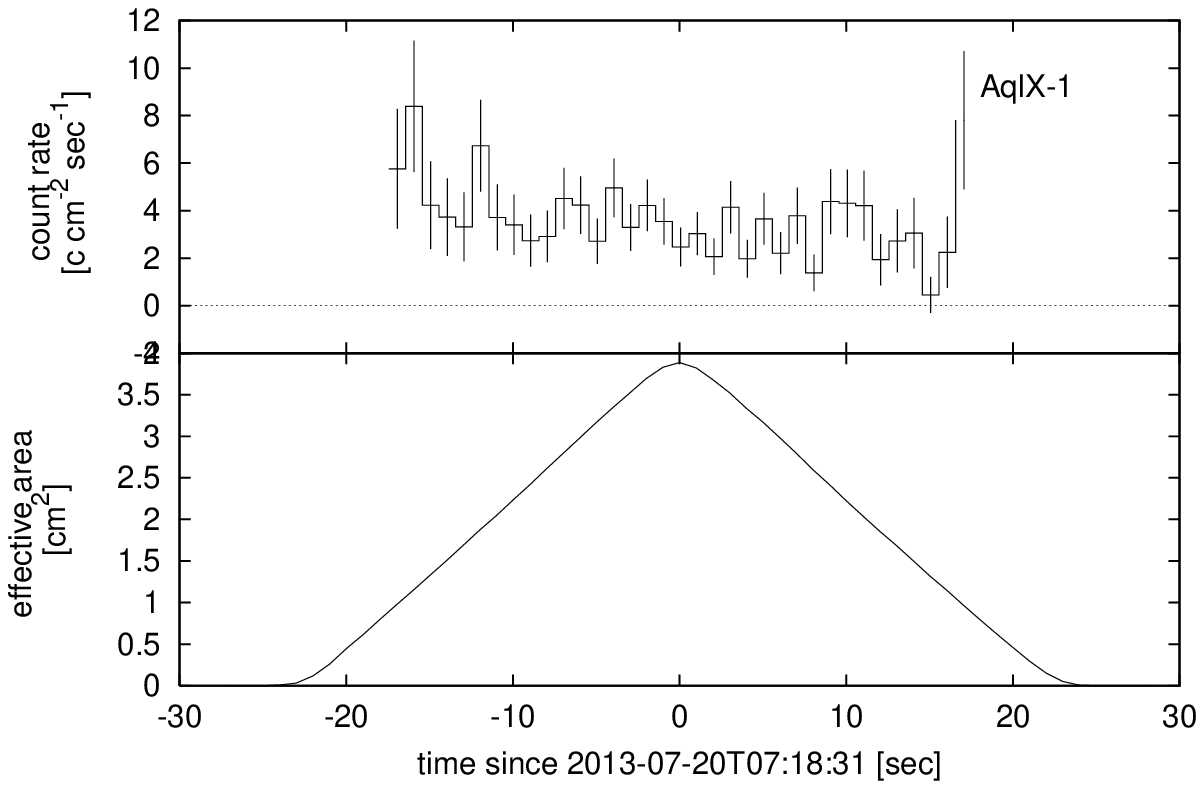}
 \end{center}
\caption{The light curves of the brightest scans of each burst
  in the 2--20 keV energy band with 1 s time resolution.
  The light curves are corrected  for the effective area,  
  whose time variation is displayed in the bottom panel of 
  each burst.
  The shaded regions represent the time intervals 
  affected by contamination sources.
  The light curve of the first burst from 4U~1850$-$086 
  (on 2014 March 10, or MJD 56726)
  is not shown 
  here, because the observed flux was too low to extract a meaningful light 
  curve with the photons observed in a scan.
  }
\label{fig:lc}
\end{figure*}

   A brief brightening with a time scale of a normal X-ray 
  burst has been commonly observed immediately before, or at 
  the onset of,  a superburst,
   and is  referred to as ``precursor''
  \citep{2002ApJ...566.1045S, 2002A&A...382..503K, 2002ApJ...577..337S, 
  2004A&A...426..257I}.
  In figure \ref{fig:lc} 
  we plotted the light curves of the bursts, 
  correcting for the effective area, 
  at the brightest scans 
  in the 2--20 keV energy band 
  with 1 s time resolution to search for potential precursors.
  The light curve of the first burst from 4U~1850$-$086 
  (on 2014 March 10, or MJD 56726)
  is not shown 
  here, because the observed flux was too low to extract a meaningful light 
  curve with the photons observed in a scan (see figure \ref{fig:blc}).

  The light curves of some sources are contaminated with an emission 
  from nearby bright sources.
  That of EXO~1745$-$248 is the worst; the entire period is 
  contaminated  with an emission from GX~3+1 and
  GX~5$-$1.  Hence, it is dropped from figure~\ref{fig:lc}.
  The light curves of SLX~1735$-$269 and SAX~J1747.0$-$2853 are
  partially affected  with contaminating sources. 
  The affected time intervals are indicated with shaded regions in 
  figure~\ref{fig:lc}.

  We find no obvious burst-like variability  (figure~\ref{fig:lc}).
  The light curve of 4U 0614+091 shows a hint of possible variation,
  and those of relatively short bursts from 4U 1820$-$30 and 
  4U 1850$-$086 show a trend of decay.
  All the other light curves are consistent with  a constant flux.

\subsection{Spectral analyses}
\label{ss:spec}
  We perform spectral analysis of the bursts, using {\tt XSPEC}, 
  to study a potential spectral evolution during a burst.
  The spectrum of each scan is fitted with a model,  
  unless the statistics of the data is too poor, in which case 
  the spectra are summed for multiple scans before fitted 
  (the rows in table~\ref{tab:spec}, in which the column MJD has a range, 
  indicate the cases of spectra with multiple scans combined).

  We adopt the blackbody model ({\tt Bbody} in {\tt XSPEC}), 
  which is characterized with a blackbody temperature and bolometric 
  flux, for the burst component. 
  The persistent emission from the sources may also have to be considered.
  For three weak sources (EXO~1745$-$248, SAX~J1828.5$-$1037, and
  4U~1850$-$086), the contribution of their persistent-emission
  component is negligible.
  However, the other sources
  (4U~0614+091, SLX~1735$-$269 SAX~J1747.0$-$2853, 4U 1820$-$30, 
  and Aql X-1) had a significant persistent emission, which is 
  incorporated in the model-fitting.
  We model the persistent emission with a power law ({\tt Powerlaw})
  or a power law with high-energy exponential-cutoff model 
  ({\tt Cutoffpl}),
  and assume that the persistent emission did not change during and 
  shortly before the burst. Thus we used a combined spectrum from scans 
  immediately before the burst to represent the persistent-emission
  spectrum. 

  The spectral parameters of the persistent-emission component of 
  each spectrum during a burst should have common value.
  A simple way to fulfill the requirement is
  using the fixed model parameters which are derived by 
  a fit of the persistent spectrum. 
  However, this method has a problem that the errors on the parameters 
  of the persistent-emission component 
  are not considered correctly at the fit of the burst spectra.

  Then we adopted a joint fit method. We jointly fitted all the spectra 
  before and during the burst for multiple scans, if available. 
  We linked the spectral parameters of the persistent-emission component,
  whereas the temperature and normalization of the blackbody 
  component for each spectrum were allowed to vary.
  The normalization of blackbody of the persistent spectrum was
  fixed to zero.
  For example, the burst from SLX~1735$-$269 on 2012 December 6 
  (MJD 56267)
  was detected for two scans, and thus we had three spectra in total;
  two spectra during
  the burst and one spectrum of the persistent emission before the
  burst.
  We jointly fitted these three spectra at a time, where the spectral
  parameters of the persistent-emission component were common to all
  three spectra.

  The column densities of the absorption are included
  for the two sources only that have a significant absorption:
  SAX J1747.0$-$2853 with $\sim10 \times 10^{22}$ cm $^{-1}$ 
  \citep{2004A&A...416..699N} and SAX J1828.5$-$1037 with
  4.1 $\times 10^{22}$ cm $^{-1}$ \citep{2008ATel.1831....1D}.
  We used {\tt Wabs} model to incorporate the column densities 
  as a fixed parameter in the model fitting. 

  Table~\ref{tab:spec} summarizes the result of the spectral fittings.
  We find that the observed peak temperatures  were typical
  of  X-ray bursts. All the sample events showed  cooling, 
  which were particularly prominent in the early phases. 
  This cooling is consistent with the difference in the light-curves 
  in the two different energy-bands;  the flux in the 4--10 keV band 
  decayed faster than  that  in the 
  2--4 keV band in  most  bursts (figure \ref{fig:blc}).
  This is the evidence that the phenomenon is not an accretion event
  but a thermonuclear burst.

\subsection{Calculation of burst parameters}
\label{ss:par}
  In table~\ref{tab:bst}, we list the parameters of the bursts.
  $F_\mathrm{obs}$ is the bolometric flux at the observed peak time 
  derived with spectral fitting (section \ref{ss:spec}). 
  $F_\mathrm{max}$, $\tau_\mathrm{LB}$, $d$, and $E_\mathrm{b}$ are 
  the possible maximum bolometric flux,
  the e-folding time, the distance to the source, and the total
  energy of the burst, respectively.
  
  For the first burst (on 2014 March 10, or MJD 56726) of 
  4U~1850$-$086, $F_\mathrm{obs}$ in the table is taken simply from 
  \citet{2011ATel.3625....1I}. 
  Similarly, $\tau_\mathrm{LB}$ of 4U 1820$-$30 and 4U 1850$-$086 (first) 
  in table~\ref{tab:bst} are taken from \citet{2011ATel.3625....1I} and 
  \citet{2014ATel.5972....1I}, 
  respectively.

  For the other sources,
  we fit the light curves of the bolometric fluxes (table~\ref{tab:spec})
  with a simple exponential function and estimate $\tau_\mathrm{LB}$ .
  Then we estimate the possible maximum bolometric flux 
  $F_\mathrm{max}$ by extrapolating the exponential function to
  the epoch of the previous scan transit.

  For SAX J1747.0$-$2853, we calculate $F_\mathrm{max}$, using
  the peak time observed by INTEGRAL \citep{2011ATel.3183....1C}
  instead of the time of the previous scan transit.
  The calculated $F_\mathrm{max}$ is
  1.9 $\times 10^{-8}$ ergs cm$^{-2}$ s$^{-1}$. 
  \citet{2011ATel.3183....1C} reported the peak flux as
  6.7 $\times 10^{-8}$ ergs cm$^{-2}$ s$^{-1}$ in 3--30 keV 
  (including persistent flux). 
  Using the temperature at the peak and assuming that the persistent 
  component was not changed, we calculate the bolometric flux at the
  peak to be 6.9 $\times 10^{-8}$ ergs cm$^{-2}$ s$^{-1}$.
  This peak flux is higher than the peak flux estimated from MAXI data. 
  The fact indicates that the initial decay was much faster than the 
  decay in late time or that there was a normal X-ray burst at the 
  beginning of the superburst.

  The burst energy $E_\mathrm{b}$ is calculated, using $F_\mathrm{obs}$,
  $\tau_\mathrm{LB}$ and $d$. Note that these $E_\mathrm{b}$ are the 
  lower limits of the true energy, because the true peak fluxes of the 
  bursts would be larger than the observed values, if the peak-flux time 
  had been out of the observed window;
  in that case, the burst energy could be as large as 
  $E_\mathrm{b} (F_\mathrm{max}/F_\mathrm{obs})$.
  The derived burst energies $E_\mathrm{b}$ are distributed around 
  10$^{41}$--10$^{42}$ erg,
  which is comparable with a typical energy of 
  intermediate-duration bursts ($10^{41}$ erg, \cite{2008A&A...484...43F})
  and superbursts ($10^{42}$ erg, \cite{2004NuPhS.132..466K}).

  The shortest e-folding time in our sample is $\sim$0.27 hours 
  from 4U 1850$-$086
  (on 2014 March 10, or MJD 56726),
  and the longest one is 5.2 hours from 4U 0614+091
  (on 2014 November 3, or MJD 56964).
  We then compare the
  e-folding times of our sample with the previous ones, if any, 
  in literature.
  The previous superburst from 4U 0614+091 
  (on 2005 March 12, or MJD 53441)
  was reported
  by \citet{2010A&A...514A..65K} and the e-folding time was 2.1 hours.
  The one from 4U 1820$-$30 
  (on 1999 September 9, or MJD 51430)
  \citep{2002ApJ...566.1045S}
  had an e-folding time of $\sim$ 1 hour. The e-folding time
  of the event from 4U~1820-30 in
  our sample (table~\ref{tab:bst}) may appear to be shorter than 
  that by a factor of two.
  However, 
  the superburst that the MAXI/GSC detected could well have 
  started roughly 0.5 hr before the scan, and in that case
  the light-curves of both the superbursts would be consistent with
  each other, as pointed out by \citet{2011ATel.3625....1I}.

   There are two bursts from 4U1850$-$086 in our sample and the 
  e-folding times of the first 
  (on 2014 March 10, or MJD 56726)
  and second 
  (on 2015 May 9, or MJD 57151)
  bursts  were 0.27  and 0.71 
  hours, respectively. A large variation in the decay time of superbursts
  from the same source has been known for at least three superburst 
  sources:
  4U 1636$-$53 \citep{2001ApJ...554L..59W},
  GX 17+2 \citep{2004A&A...426..257I}, and 
  Ser X-1 \citep{2002A&A...382..174C,2009ATel.2140....1K}.
  Our discovery has added another example to confirm the large variation 
  in the decay time in long-lasting X-ray bursts.
  We discuss the classification of the sample events by the duration 
  in section \ref{ss:overall}.

\subsection{First type-I bursts after the long bursts}
  Burst quenching is a common phenomenon after a superburst 
  \citep{2000A&A...357L..21C,2002A&A...382..503K}.
  Table \ref{tab:bst} lists the time of the first X-ray bursts after 
  the long bursts that we detected. 

  \citet{2011ATel.3217....1L} reported that a normal X-ray
  burst from SAX~J1747.0$-$2853 
  (on 2011 February 13, or MJD 55605)
  was observed 25 days after the
  superburst.
  Normal X-ray bursts from 4U~1820$-$30 and 
  Aql~X-1 were observed by MAXI 
  1549 and 389 days after the long bursts, respectively.
  The first X-ray burst from 4U~1850$-$086 after the long burst
  on 2014 March 10 (MJD 56726) was observed on 2015 May 9 (MJD 57151,
  425 days after the long burst), which is the second event in our 
  sample.
  After the second event,
  another X-ray burst with a longer duration than the scan 
  transit ($\sim$40 sec) on 2015 November 11 (MJD 57337) occurred from the
  same source (see section \ref{ss:type}). Thus, the quenching time
  of the second event was 186 days.
  No X-ray burst was detected by MAXI after the long bursts for
  the other sources (4U~0614+091, SLX~1735$-$269, EXO~1745$-$248,
  and SAX~J1828.5$-$1037).

\subsection{Persistent emissions of the sources}
\label{ss:persistent}
  MAXI can monitor the persistent fluxes of the burst sources,
  and therefore MAXI is an optimum instrument for studying the relation
  between the persistent emissions and superbursts and the potential 
  change of the persistent flux after  a superburst.
  Using the MAXI light curves of one-day bin in the public data archive, 
  we calculate ten-day 
  average fluxes in the 2--20 keV band (10 days -- 1 day) before and 
  (1 day -- 10 days) after each burst.%
  \footnote{
    For 4U~0614+091, we use the data of 30--25 days and 1 day before,
    and 2--10 days after the burst,
    because the public light curve of 4U 0614+091 is not available
    from 24--2 days before and 1 day after the burst.
  }
  The photon count rates  are converted to the energy fluxes, 
  assuming the crab-like spectrum.
  The results are  given in the columns $F_\mathrm{before}$ and 
  $F_\mathrm{after}$ in table~\ref{tab:pers}.

\begin{figure*}
 \begin{center}
   \includegraphics[width=8cm]{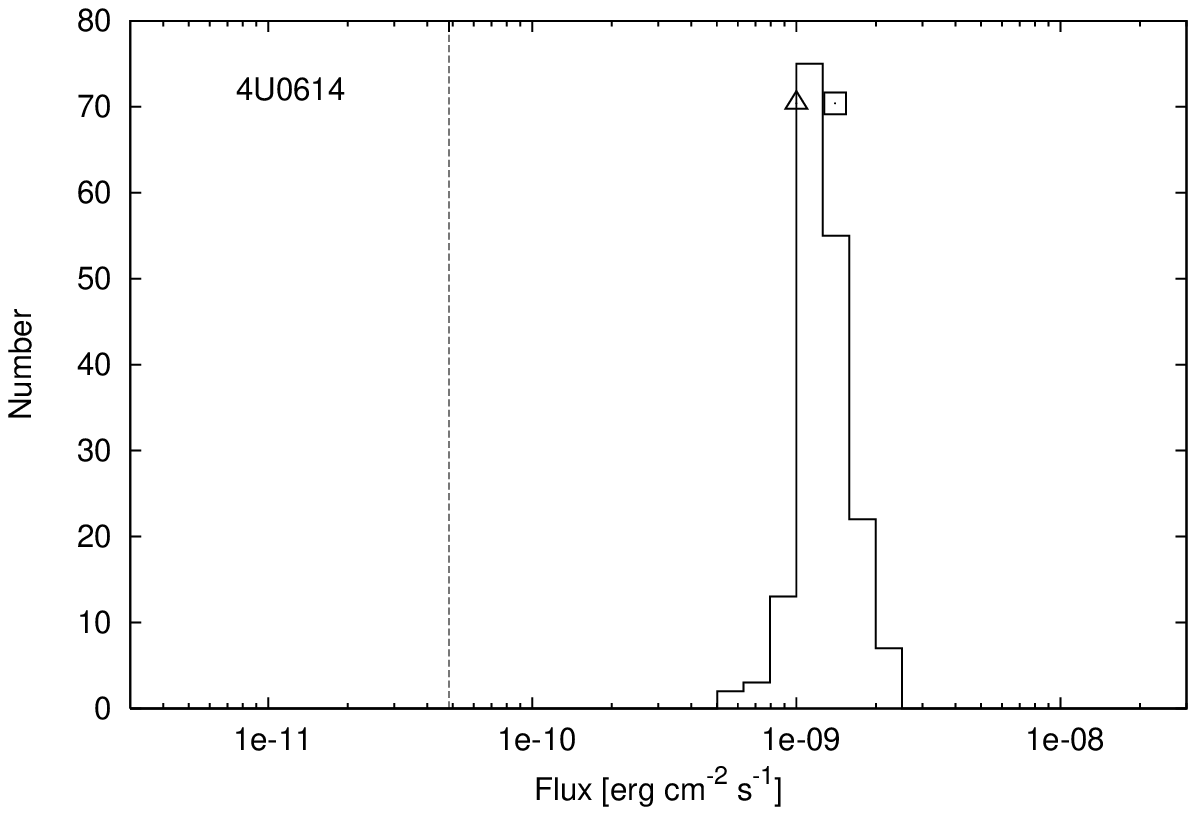}
   \includegraphics[width=8cm]{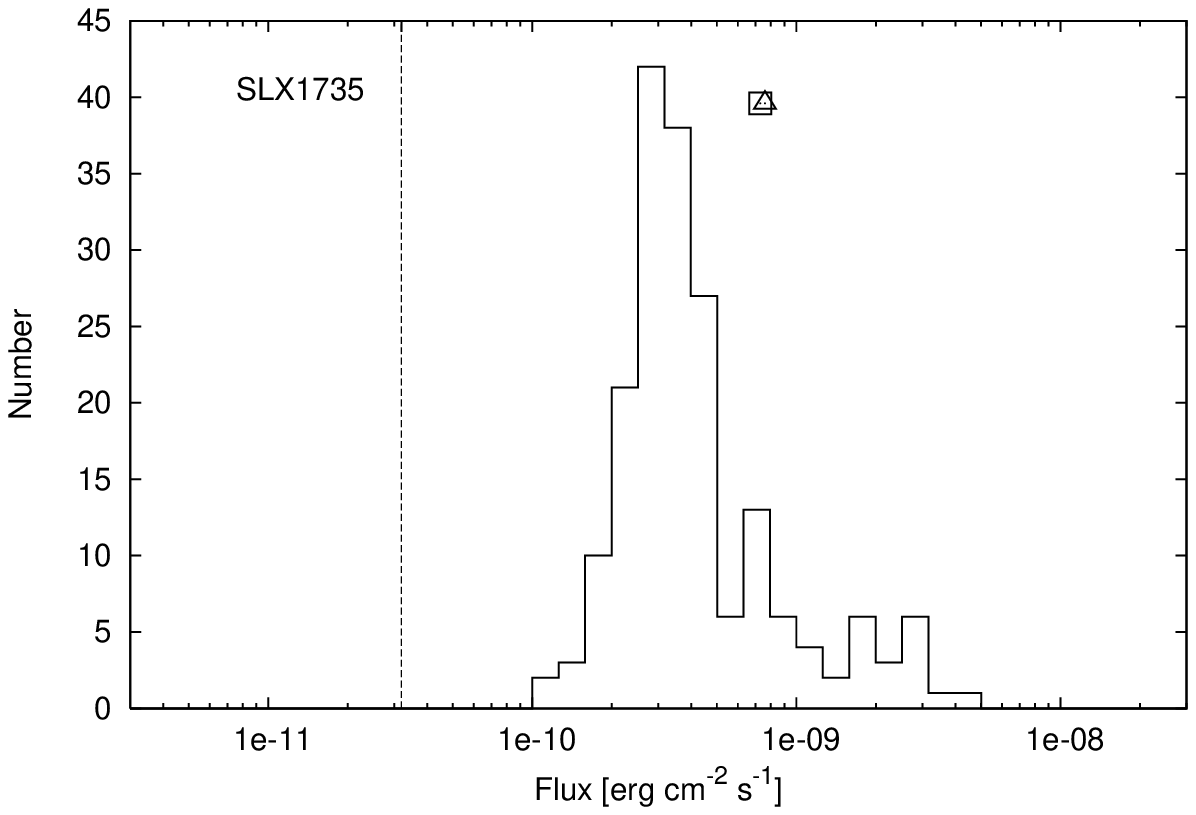}
   \includegraphics[width=8cm]{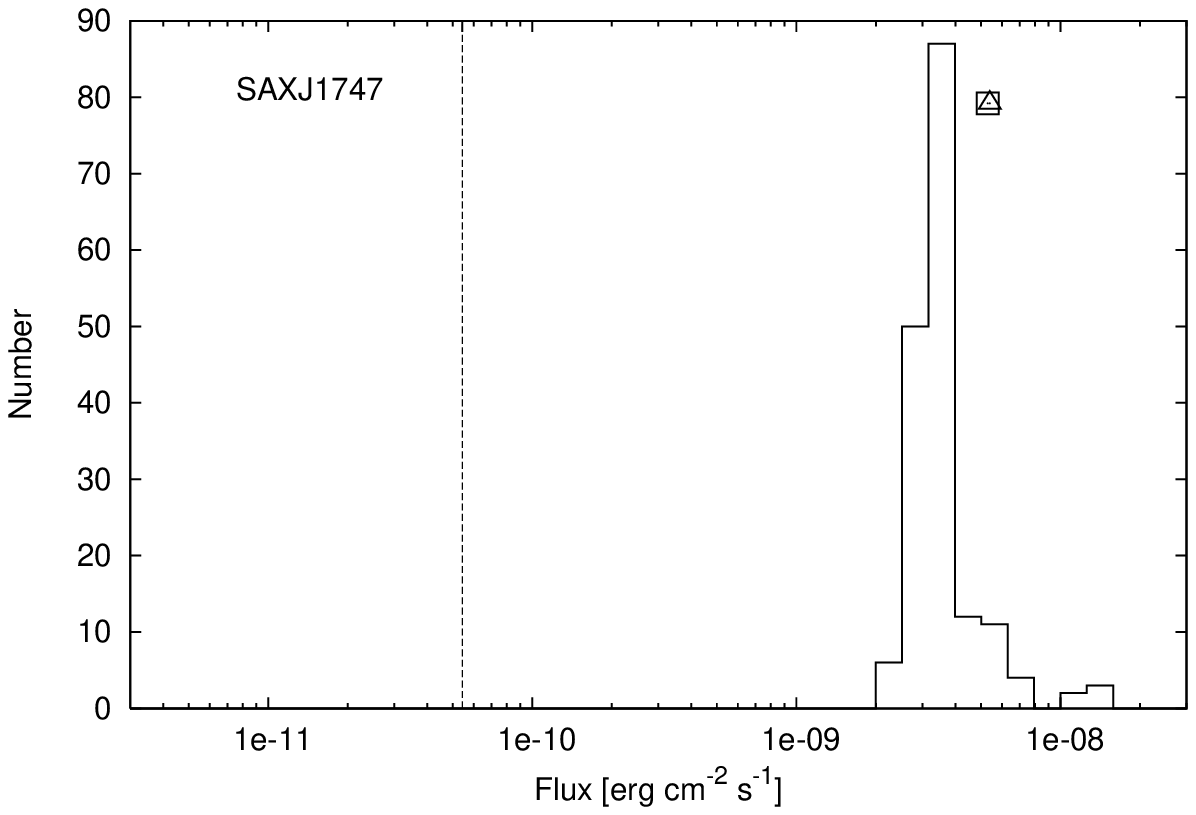}
   \includegraphics[width=8cm]{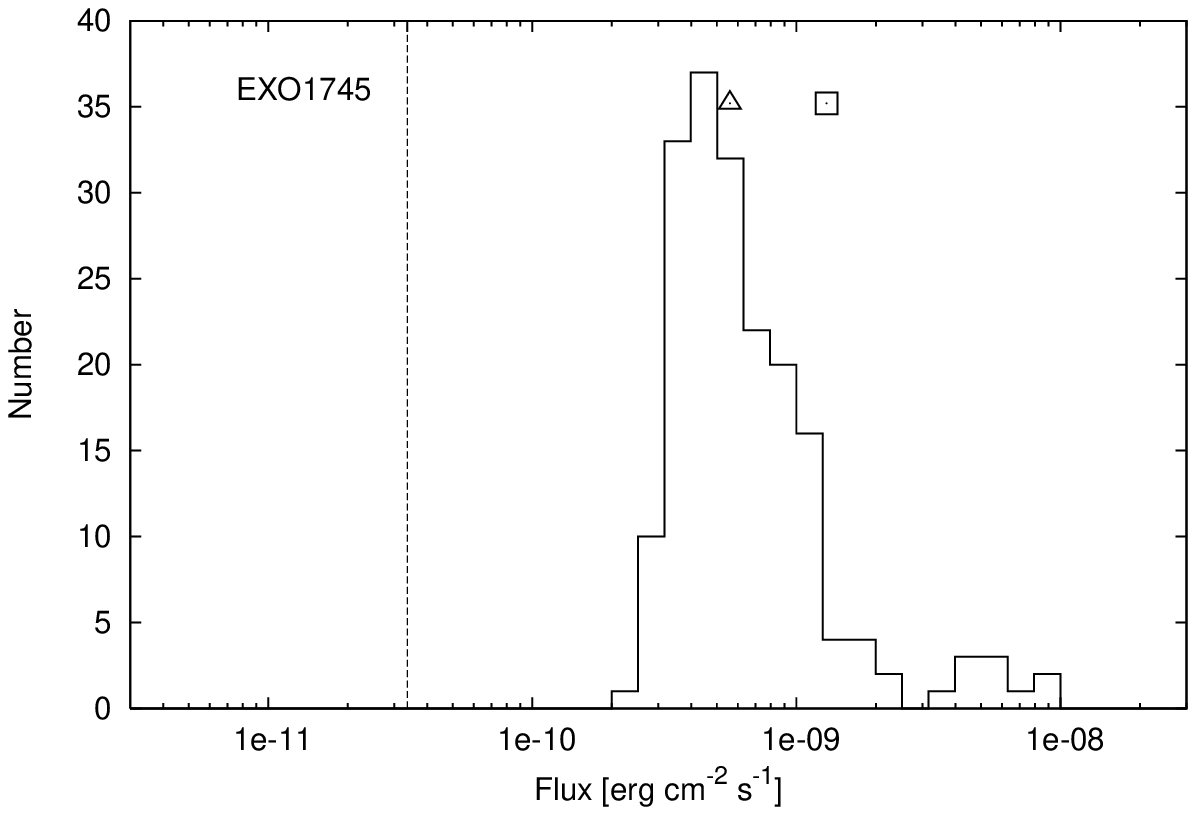}
   \includegraphics[width=8cm]{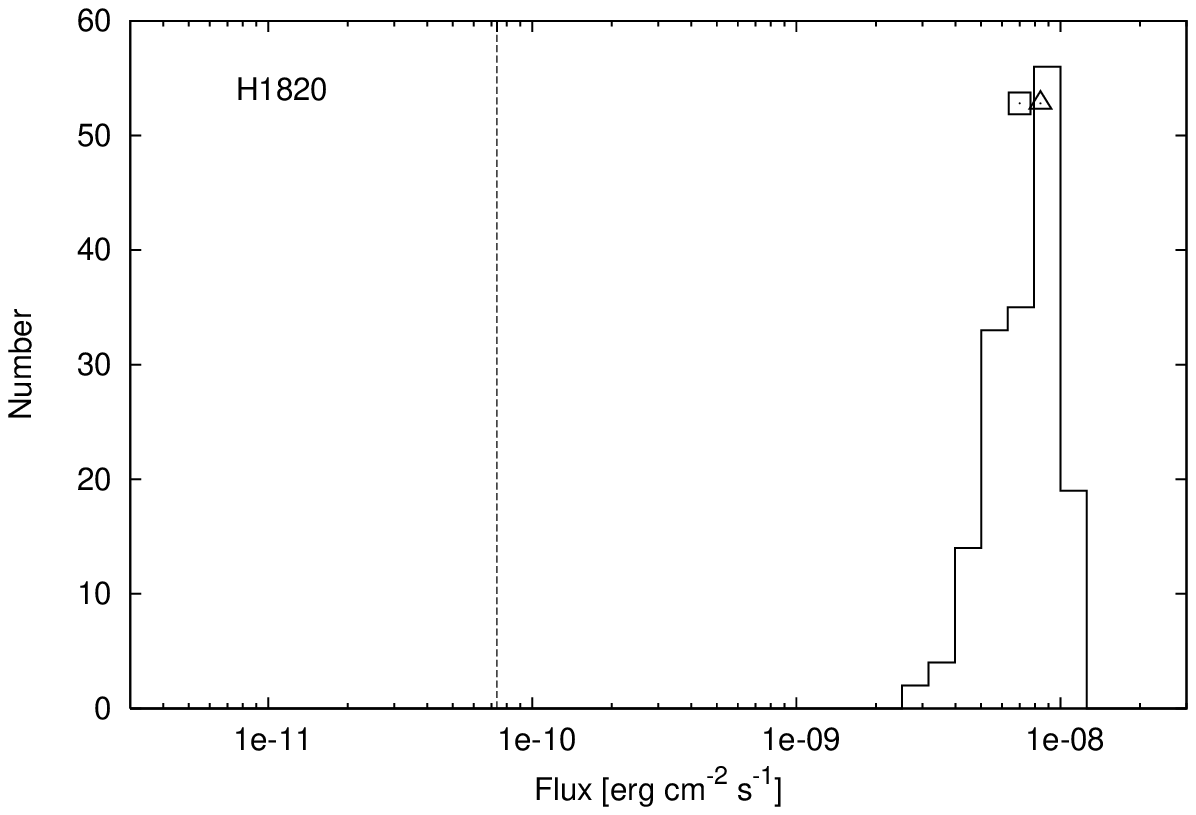}
   \includegraphics[width=8cm]{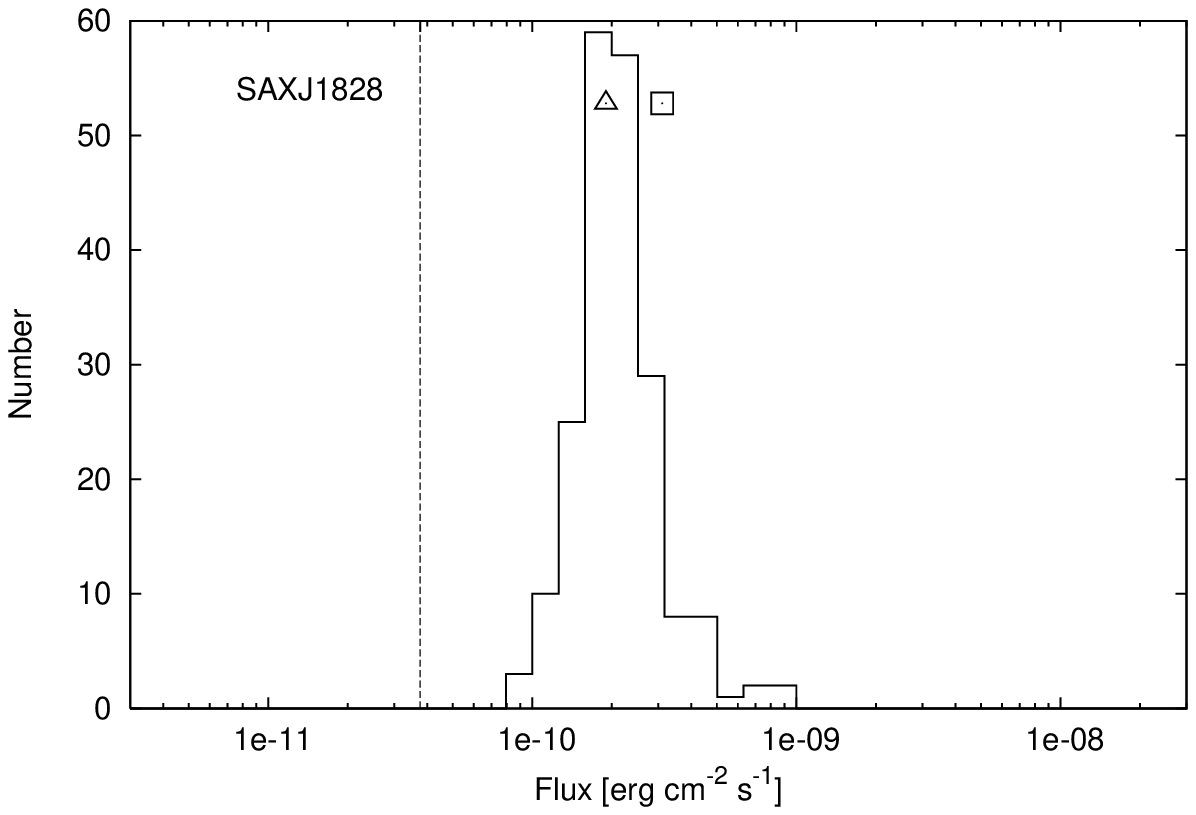}
   \includegraphics[width=8cm]{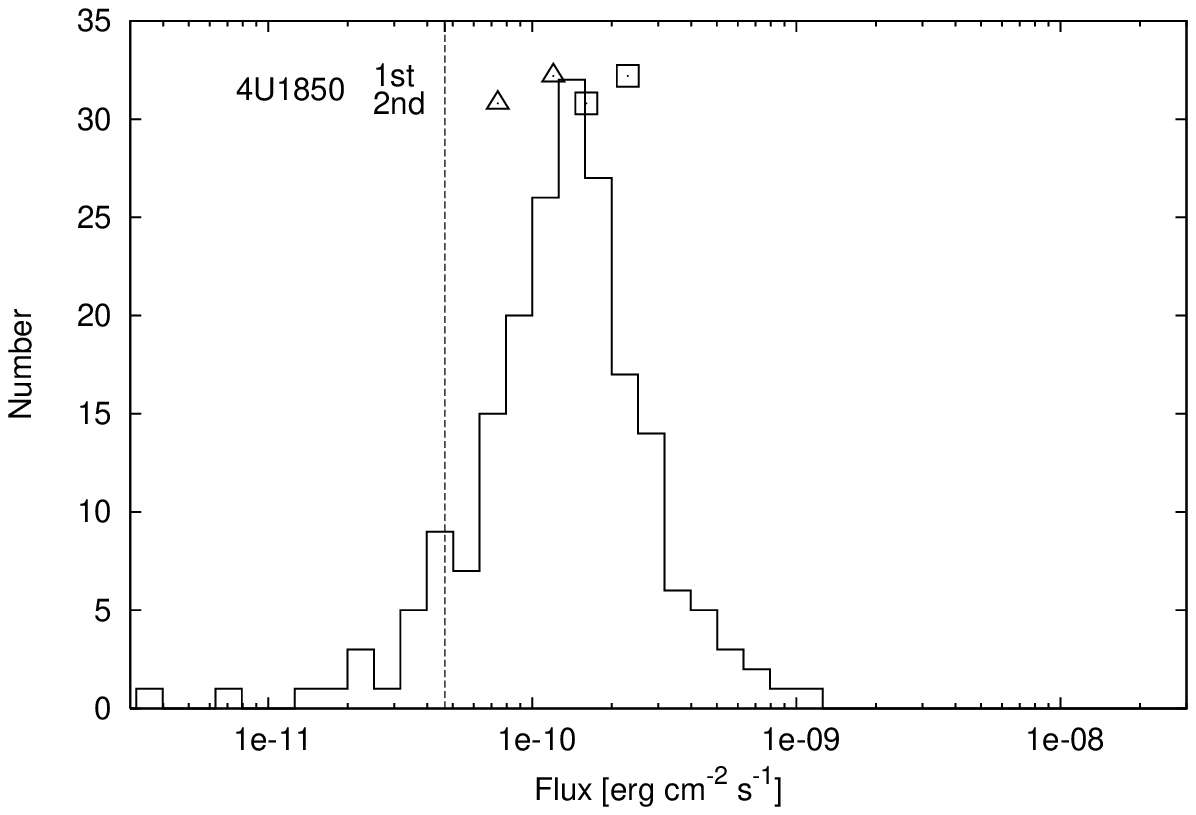}
   \includegraphics[width=8cm]{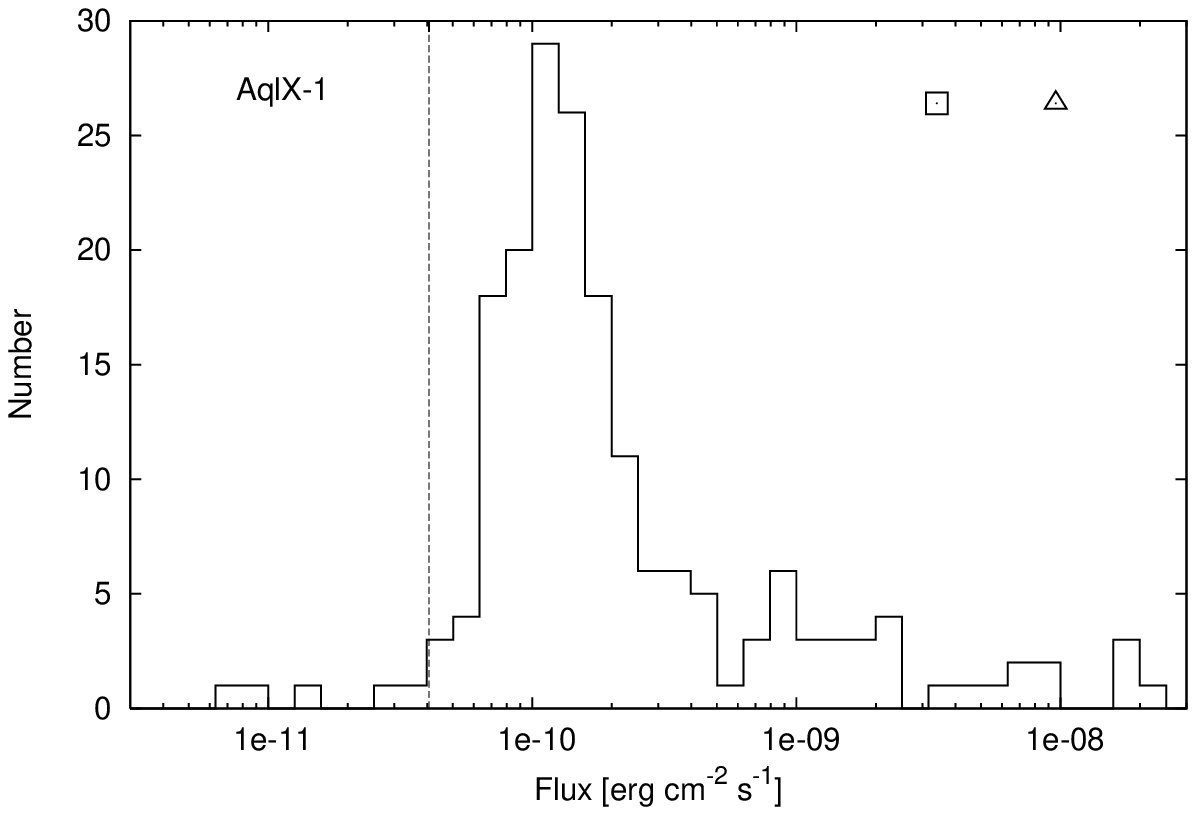}
 \end{center}
\caption{The persistent fluxes of each source before (triangles) and 
         after (squares) the bursts. The histograms are the distributions
         of ten-day average flux of the sources.
         The vertical dashed-lines indicate the typical errors
         on the ten-day average fluxes of the sources,
         which correspond to the 1 sigma detection limit for ten days.
         }
\label{fig:persistent}
\end{figure*}

  We  find some significant changes in the flux after the bursts
  for some sources.
  In order to compare the changes of the persistent fluxes 
  after the bursts with the intrinsic variations,
  we plotted in figure~\ref{fig:persistent} the distribution 
  of the ten-day average fluxes of each source in the same energy band, 
  using all the available MAXI data. 

  An increase in the persistent flux is apparent in 
  EXO~1745$-$248, and is marginally confirmed in SAX~J1828.5$-$1037,
  4U~0614+091, and the first sample of 4U~1850$-$086
  (on 2014 March 10, or MJD 56726).
  The burst from Aql X-1 occurred during the decay part of an 
  outburst while the persistent emission was continuously decreasing.
  Therefore, the persistent flux after the superburst 
  is expected to be significantly lower than that before the burst.

\section{Discussion and conclusion}
\label{ss:dis}

\subsection{Discussion about overall properties}
\label{ss:overall}

  Before the MAXI era, most of the known superburst
  sources  were persistent and bright.
  \citet{2004NuPhS.132..466K} summarized the persistent luminosities of
  superbursters, and concluded that all of them were at or above 10\% of
  Eddington luminosity.
  In contrast, \citet{2008A&A...484...43F} showed that
  intermediate-duration bursts could occur at $\leq$ 1\% Eddington 
  luminosity.

\begin{figure}
 \begin{center}
   \includegraphics[width=8cm]{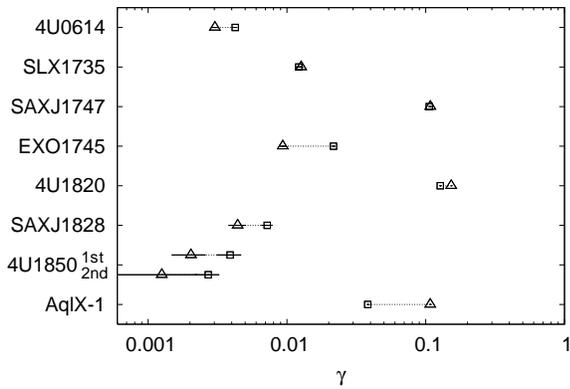}
 \end{center}
  \caption{The persistent fluxes before (triangles) and after 
  (squares) the long bursts, 
  normalized  to the peak fluxes of normal bursts $\gamma$.
  For 4U~1850$-$086, there are two pairs of points, corresponding
  to the two bursts.}
\label{fig:pn}
\end{figure}

  Figure~\ref{fig:pn} shows the ratio $\gamma$ of the persistent
  fluxes of each source before and after the superburst to 
  the peak flux of the normal bursts from the same source.
  The peak flux of normal bursts of a burster is usually its Eddington
  limit or close. Accordingly, $\gamma$ here is a good approximation 
  to the Eddington ratio of the  
  source fluxes
  \citep{1993SSRv...62..223L}.
  4U~1820$-$30 has the largest $\gamma$ in our sample,
  which is consistent with that listed in 
  \citet{2004NuPhS.132..466K}.
  Six of the other sources are transients (table~\ref{tab:pers}).
  The calculated ratios of SAX~J1747.0$-$2853 and Aql~X-1 are around 0.1 
  and the superbursts of these sources occurred during outbursts.
  The ratios are $\lesssim$ 0.01 for 
  six out of nine in our MAXI sample.
  Among them, two sources, EXO 1745$-$248 and SAX J1828.5$-$1037, are 
  transient 
  and they might have been in quiescence at the time of the superbursts.
  The other four of the six are ultracompact X-ray binaries (UCXBs) or 
  candidate UCXBs \citep{2007A&A...465..953I},
  which are known to have persistent luminosity of a small Eddington ratio.

  We should note that a selection bias may be significant;
  it is easier with MAXI to find superbursts
  from sources with weaker persistent emissions, because burst
  signals over fluctuations
  of the persistent component are more prominent in fainter sources 
  than brighter ones. 
  We need further investigations to find whether
  superbursts from bright sources exist in the MAXI data or not.

\begin{figure}
 \begin{center}
   \includegraphics[width=8cm]{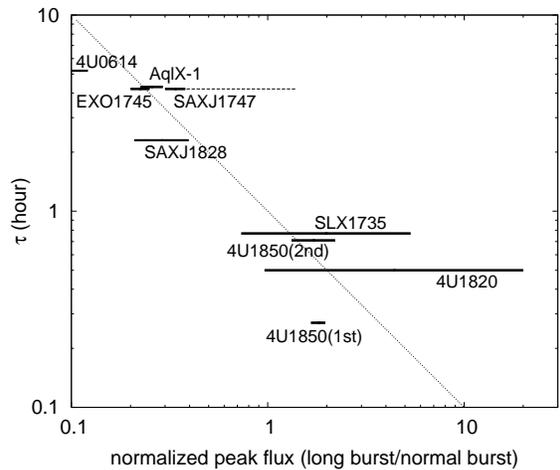}
 \end{center}
  \caption{A scatter plot of the fluxes of the long bursts normalized 
   to those of normal bursts and $\tau_\mathrm{LB}$ of the long bursts.
  The lower ends of the fluxes are the observed ones ($F_\mathrm{obs}$),
  and the upper ends are possible maximum fluxes ($F_\mathrm{max}$).
  For the first event of 4U 1850$-$086, we simply plot the error of
  the observed flux.
  The dashed line of the SAX~J1747.0$-$2853 shows $F_\mathrm{max}$
  observed by INTEGRAL.
  The  relation [normalized flux $=1/\tau$] is shown with a dotted line.
  }
\label{fig:sn}
\end{figure}

  Next, we plot in figure~\ref{fig:sn} 
  the e-folding decay time $\tau_\mathrm{LB}$ of the burst versus the 
  normalized peak flux, 
  which is the peak flux of our 
  sample ($F_\mathrm{obs}$ or $F_\mathrm{max}$ in table~\ref{tab:bst}) 
  normalized to the peak flux of normal bursts ($F_\mathrm{b}$
  in table~\ref{tab:pers}). 
  Four bursts
  (from SLX~1735$-$269, 4U~1820$-$30, and two bursts from 4U~1850$-$086)
  showed an e-folding decay time $\tau_\mathrm{LB}$ shorter than one 
  hour and had a relatively high peak flux.
  4U~0614+091, SLX~1735$-$269, 4U~1820$-$30 and 4U~1850$-$086 are
  (or are suppose to be) UCXBs, from which 
  intermediate-duration bursts are often observed.
  The observed durations and fluxes of the bursts from these sources
  except 4U~0614+091 are consistent with intermediate-duration bursts.
  Therefore, we argue that the four long bursts 
  with an e-folding decay time $\tau_\mathrm{LB}$ shorter than one hour
  which MAXI detected from 
  SLX 1735-269, 4U 1820-30 and 4U 1850-086 should be classified as 
  intermediate bursts. 
  Notably, the burst from SLX 1735-269 is the longest intermediate 
  burst ever found. Consequently, the other five events are superbursts.

  The derived peak fluxes have a large ambiguity due to 
  the unknown delay from the peak time of the burst to the scan time. 
  Nevertheless, the peak flux and 
  $\tau_\mathrm{LB}$ seem to show a significant anti-correlation.
  In other words, brighter bursts decay faster.  Consequently, the total
  emitting energy of the bursts clusters around 10$^{41}$--10$^{42}$ erg.
  Given that the data contain two classes of intermediate-duration burst 
  and superburst,
  it appears that bursts with a wide range of durations follow the same
  relation. The suggested anti-correlation between the peak flux
  and the e-folding time does not hold for normal bursts, because the
  normalized peak fluxes of normal bursts are by definition equal to, or
  smaller than, unity, whereas the durations of normal bursts are more than 
  an order of magnitude shorter than an hour.

  We calculate the parameters of nuclear reactions, using
  the cooling model given by \citet{2004ApJ...603L..37C}. 
  The model parameters are the ignition column depth $y_{12}$
  in units of 10$^{12}$ g cm$^{-2}$ 
  and energy release per unit mass $E_{17}$ in units of 
  10$^{17}$ erg g$^{-1}$. 
  The model approximates the light curves of the bursts with
  a broken power-law function, which has the decay indices of
  $-0.2$ and $-4/3$ in the early and late phases, respectively.
  The parameter $t_\mathrm{cool}$, indicating the transition time 
  between the two indices, is  proportional to $y_{12}^{3/4}$.
  Therefore, a burst with a larger ignition column depth $y_{12}$
  stays bright for a longer period.
  The luminosity $L$ of a burst is a direct function of the energy 
  release $E_{17}$: $L \propto E_{17}^{7/4}$ and 
  $L \propto E_{17}^{1/2}$ in the early and late phases, respectively.
  Thus, we can obtain these parameters by
  fitting the light curves of the bursts. 

  This analysis requires good statistics.  Hence,
  we used the events with
  more than two observed scans only:  4U~0614+091, SAX J1747.0$-$2853, 
  EXO 1745$-$248, SAX J1828.5$-$1037,
  and Aql X-1. For EXO 1745$-$248, we  considered the cases of two  
  possible distances.
  The uncertainties of the burst peak time (and the flux at that time)
  are the major cause of the systematic error,
  because the peak-flux time may be out of the observed window
  as discussed in section \ref{ss:par}. 
  Note that the maximum values of the peak fluxes are listed in
  the column  $F_\mathrm{max}$ in table~\ref{tab:bst} and the minimum
  values are equal to $F_\mathrm{obs}$ in the table.

\begin{figure}
 \begin{center}
   \includegraphics[width=8cm]{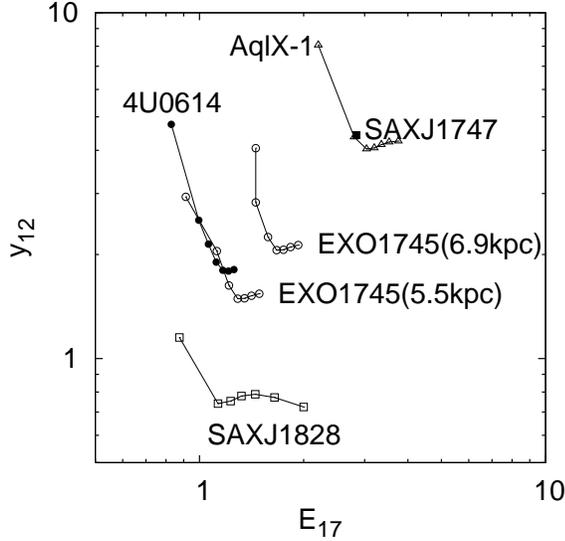}
 \end{center}
  \caption{A scatter plot of the released energy 
  $E_{17}$ (10$^{17}$ erg g$^{-1}$) and 
  the ignition column depth $y_{12}$ (10$^{12}$ g cm$^{-2}$).
   Each of seven points per event, connected with a line, corresponds 
   to the result of  fitting with  a different assumed peak time,  
   $\delta t$ days before the first observation 
  ($\delta t$ = 0.001, 0.01, 0.02, 0.03, 0.04, 0.05, or 0.06).
  The results of the same source are plotted with the same symbol 
  and are connected with a solid line. 
  For EXO~1745$-$248,  two different cases for different distances 
  are plotted.
  In the case of the burst from SAX J1747.0$-$2853, the burst peak was 
  observed by INTEGRAL, 
  and hence we use the time of it (a single point in the figure).
  }
\label{fig:cool}
\end{figure}

  We perform calculation of the model parameters for seven possible peak
  times for a burst 
  with model fitting, except for SAX J1747.0$-$2853, 
  where we simply accept the burst peak-time observed with 
  INTEGRAL. Figure~\ref{fig:cool} shows the results.
  Compared with the works in literature for the past superbursts
  (for example, \cite{2006ApJ...646..429C,2008A&A...479..177K}),
  our results are in the typical range of the parameters.

  Burst quenching has been studied systematically by 
  \citet{2004ApJ...603L..37C} and \citet{2012ApJ...752..150K}.
  There is only an event with a short ($<$100 days) quenching time 
  in our sample (table~\ref{tab:bst}).
  A normal X-ray burst from SAX~J1747.0$-$2853 was observed 25 days 
  after the superburst
  \citep{2011ATel.3217....1L}.
  The model by \citet{2004ApJ...603L..37C} gives an estimation of
  the quenching time. Using $y_{12}$, $E_{17}$
  (figure \ref{fig:cool}) and $\gamma$ (figure \ref{fig:pn}),
  we calculated the quenching time to be 26 days, which is consistent with
  the observed one.
  Good agreement between the predicted and the observed values of
  quenching time have been reported for superbursts (for example
  from 4U~1636$-$536 on 1999 June 26 (MJD 51324) 
  \cite{2004AIPC..714..257K};
  from GX~17+2 on 1996 September 14 (MJD 50340) and 
  on 1999 October 1 (MJD 51452) \cite{2004A&A...426..257I};
  from 4U~0614+091 on 2005 March 12 (MJD 53441) 
  \cite{2010A&A...514A..65K}). Our result adds another
  example to that, and confirms the relation.

\subsection{Discussion on individual sources}

\subsubsection{4U 0614+091}
 \label{ss:0614} 
   A superburst from 4U 0614+091 was observed on 2005 March 12 
   (MJD 53441) with RXTE/ASM before the MAXI era.  
   \citet{2010A&A...514A..65K} reported that the source had experienced  
   brightening in the hard X-ray (15--50 keV) data
  before the superburst. Then, we look through the light curve of
  Swift/BAT transient monitor results provided by the Swift/BAT team
  \citep{2013ApJS..209...14K}, and  find that the superburst observed
  by MAXI also occurred at the time of the high persistent flux.
  These facts suggest that some special accretion condition 
  may induce superbursts.

  \citet{2010A&A...514A..65K} pointed out that this source showed 
  flaring and quiet periods in the persistent emission.
  They found that an intermediate-duration burst occurred during the 
  calm period and that all the other bursts including the superburst 
  occurred during the flaring period.
  We have found the same trend in the MAXI data; the 
  superburst observed by MAXI also occurred during the flaring period.

\subsubsection{Low accretion rate  in EXO 1745$-$248 
      and SAX J1828.5$-$1037 before the superbursts}
  We find that EXO 1745$-$248 and SAX J1828.5$-$1037 had
  a very weak persistent emission before the superbursts.
  It is unclear how these superbursts were ignited during the
  periods of low accretion rate.
   Indeed, \citet{2012MNRAS.426..927A} has argued that the low 
  accretion rate of EXO 1745$-$248 before the superburst makes the
  ignition difficult.
  Another case of a superburst during the period of low accretion was
  from so-called ``burst-only source'', SAX J1828.5$-$1037
  \citep{2002A&A...392..885C,2002A&A...392..931C}.
   \citet{2009ApJ...699.1144C} estimated the 0.5--10 keV unabsorbed 
  flux in  quiescence  to be 
  1.5 $\times$ 10$^{-13}$ erg cm$^{-2}$ s$^{-1}$.
  \citet{2004MNRAS.351...31H} reported that the source  had been observed
  2001 March and 2002 September, but had not been detected  in the latter 
  observation with the upper limit  of the flux of 
  2 $\times$ 10$^{-14}$ erg cm$^{-2}$ s$^{-1}$ in the 2--10 keV band.

  At least two outbursts have been observed from SAX J1828.5$-$1037 
  so far, in 2001 \citep{2004MNRAS.351...31H} and  
  in 2008 \citep{2008ATel.1831....1D}.
  The flux level during the latter outburst was at about
  10$^{-12}$ erg cm$^{-2}$ s$^{-1}$.
  The distribution of 10-day average fluxes of this source observed in
  2011 by MAXI/GSC (figure \ref{fig:persistent}) clusters around 
  10$^{-10}$ erg cm$^{-2}$ s$^{-1}$, and the fluxes are higher than
  those of above-mentioned outburst.
  Note that the observed flux in figure \ref{fig:persistent} 
  is higher than the real one due to the contamination of the Galactic 
  ridge emission. MAXI can not detect an outburst with the flux of
  10$^{-12}$ erg cm$^{-2}$ s$^{-1}$.
  On the other hand, the peak flux of normal X-ray burst from this 
  source was 4.3 $\pm$ 1.6 $\times$ 10$^{-8}$ erg cm$^{-2}$ s$^{-1}$
  (table~\ref{tab:pers}).
  An outburst with the flux larger than 1/100 of that at the peak of 
  an X-ray burst should be detected by MAXI/GSC, if it occurs.

  In both the cases of EXO 1745$-$248 and SAX J1828.5$-$1037,
  an increase of the persistent flux was observed
  (figure \ref{fig:persistent}). 
  \citet{2012PASJ...64...91S} has shown that the superburst of 
  EXO 1745$-$248 was followed by a short outburst
  for about five days in the light curve EXO~1745$-$248.
  They discussed
  the possibilities that the outburst was induced by the superburst.
  The increase of the persistent flux of SAX J1828.5$-$1037 may be
  another example of an outburst following a superburst.

\subsubsection{Correlation between accretion rate and burst type  in
4U 1850$-$086}
\label{ss:type}

\begin{figure*}
 \begin{center}
   \includegraphics[width=16cm]{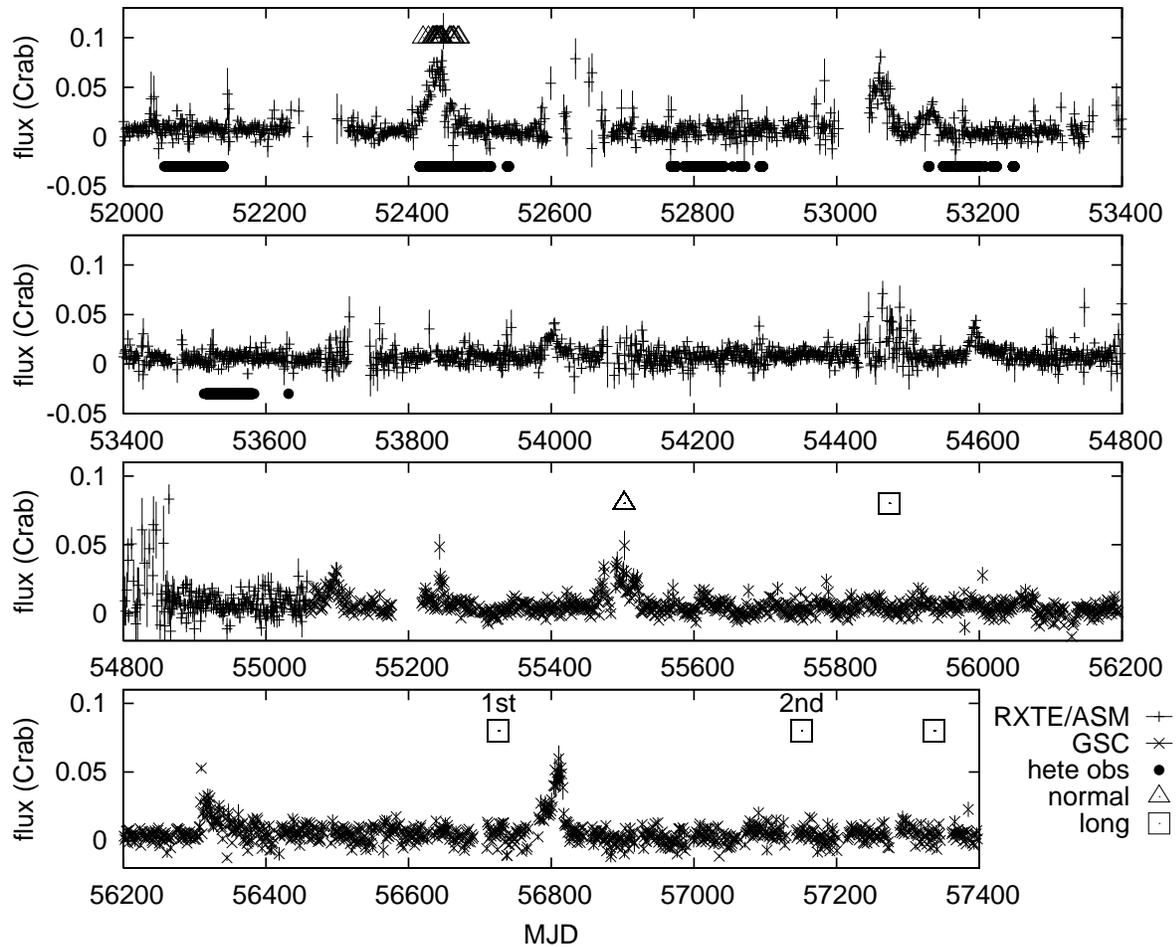}
 \end{center}
\caption{The persistent fluxes of 4U 1850$-$086 and the epochs of bursts.
      The persistent fluxes observed by RXTE/ASM (plus) and MAXI/GSC 
      (cross) are plotted.
      The circles show the  periods when the source  was in 
      the HETE-2/WXM FoV.
      The triangles and squares show the time of normal and long bursts, 
      respectively.
      }
\label{fig:1850}
\end{figure*}

  Although RXTE observed 1187 X-ray bursts in more than 10 years of 
  observations, none was observed from 4U~1850$-$086
  \citep{2008ApJS..179..360G}.
  In contrast, HETE-2 observed 28 normal bursts from 4U~1850$-$086 
  in 2002 \citep{2006NCimB.121.1593S}.
  In general, the activities of normal X-ray bursts are thought to be 
  related to the persistent luminosities.
  Then, we plot the persistent flux observed by RXTE/ASM%
  \footnote{quick-look results provided by the ASM/RXTE team:
  http://xte.mit.edu/ASM\_lc.html}
  or MAXI and the time of  normal bursts in figure \ref{fig:1850}.
  The source was in the HETE-2/WXM field of view (FoV) 
  in May--September from 2001 to 2005,
  because HETE-2 pointed toward the anti-solar direction.
  The  periods when the source  was in the HETE-2 FoV  are
  indicated in figure~\ref{fig:1850}.

  MAXI observed one normal burst and four long bursts from this source.
  Two of the long bursts are listed in table~\ref{tab:bst}.
  The durations of the others (on 2011 November 9, MJD 55874 
  and on 2015 November 11, MJD 57337)  were longer than 
  those of the scans (75 s and 40 s, respectively).
  However, the flux dropped back to the background level in the next 
  scan.
  Therefore, the durations of these bursts are unclear,
  and we have excluded them
  from the sample in this paper.

  Figure \ref{fig:1850} shows a possible correlation between the 
  accretion rate and the burst type. All the normal bursts  were observed
  when the persistent flux  was relatively high,  whereas the long bursts 
  occurred when the persistent flux  was low.
  The average persistent flux observed by ASM for 28 normal X-ray 
  bursts was 50 mCrab with the standard deviation of 25 mCrab.
  The source was not significantly detected just before the four 
  long bursts and 1-sigma flux upper-limits in these observations were 
  about 5 mCrab.
  A trend that the burst type changes in the phase of low persistent
  flux has been reported for intermediate-duration bursts 
  from 4U 0614+091 \citep[and references therein]{2010A&A...514A..65K}.
  Our result is the first indication that the trend may extend to
  long X-ray bursts from 4U 1850$-$086, although the statistics of
  the sample is still limited to derive definite conclusion.

\subsection{Conclusions}
  We studied the properties of nine long X-ray bursts from eight sources
  observed with MAXI in 2009--2015.
  We found that six out of the eight sources were transient and 
  five out of the nine bursts occurred when the persistent luminosities 
  were lower than 1\% of the Eddington luminosity.
  These trends are contrastive to the superbursts observed before the
  MAXI era, which originated from bright persistent sources.
  There is a negative correlation between the durations 
  and the peak fluxes of our sample.
  The total released energy of them clusters between 10$^{41}$--10$^{42}$ 
  erg.
  Using the light curves  with RXTE/ASM and MAXI,
  we compared the flux levels of persistent emissions 
  at the time of ``normal'' and long bursts from 4U 1850$-$086.
  We found that the four long bursts occurred  in the phase of 
  relatively low persistent flux,
   whereas normal X-ray bursts  occurred during  outbursts.
%


\begin{longtable}{*{5}{l}}
  \caption{The list of the bursts studied in this paper.}
    \label{tab:lst}
      \hline
      Name & date & MJD & inst. & ref.\\ 
      \hline
\endfirsthead
      \hline
      Name & date & MJD & inst. & ref.\\ 
      \hline
\endhead
  \hline
\endfoot
  \hline
  \multicolumn{5}{l}{
    References are 
    (A) \cite{2014ATel.6668....1S}, 
    (B) \cite{2012ATel.4622....1N}, 
    (C) \cite{2011ATel.3183....1C}, 
    (D) \cite{2011ATel.3729....1M}, 
    } \\
  \multicolumn{5}{l}{
    (E) \cite{2012PASJ...64...91S}, 
    (F) \cite{2012MNRAS.426..927A}, 
    (G) \cite{2011ATel.3625....1I},
    (H) \cite{2011ATel.3760....1A},
    } \\
  \multicolumn{5}{l}{
    (I) \cite{2014ATel.5972....1I},
    (J) \cite{2014ATel.5978....1S},
    (K) \cite{2015ATel.7500....1N}.
    } \\
  \hline
\endlastfoot
      4U 0614+091         & 2014-11-03 & 56964 & & A \\
      SLX 1735$-$269      & 2012-12-06 & 56267 & & B \\
      SAX J1747.0$-$2853  & 2011-02-13 & 55605 & INTEGRAL & C \\
      EXO 1745$-$248      & 2011-10-24 & 55858 & & D, E, F \\
      4U 1820$-$30        & 2010-03-17 & 55272 & RXTE & G \\
      SAX J1828.5$-$1037  & 2011-11-12 & 55877 & & H \\
      4U 1850$-$086 (1st) & 2014-03-10 & 56726 & Swift & I, J \\
      \hspace{13ex} (2nd) & 2015-05-09 & 57151 & & K \\
      Aql X-1             & 2013-07-20 & 56493 & & --- \\
\end{longtable}

\begin{longtable}{*{6}{l}}
  \caption{The properties of the persistent emissions and normal bursts
  of the bursters.}
  \label{tab:pers}
      \hline
     Name & transient 
      & $F_{\mathrm b}$ \footnotemark[$*$] 
      & $F_\mathrm{before}$ \footnotemark[$\dagger$] 
      & $F_\mathrm{after}$  \footnotemark[$\dagger$] 
      & ref. \footnotemark[$\ddagger$] \\ 
        &     
      & ($10^{-8}$ ergs cm$^{-2}$ s$^{-1}$)
      & \multicolumn{2}{l}{($10^{-11}$ ergs cm$^{-2}$ s$^{-1}$)} & \\
      \hline
\endfirsthead
      \hline
     Name & transient 
      & $F_{\mathrm b}$     
      & $F_\mathrm{before}$ 
      & $F_\mathrm{after}$  
      & ref. \\ 
        &     
      & ($10^{-8}$ ergs cm$^{-2}$ s$^{-1}$)
      & \multicolumn{2}{l}{($10^{-11}$ ergs cm$^{-2}$ s$^{-1}$)} & \\
      \hline
\endhead
  \hline
\endfoot
  \hline
      \multicolumn{6}{l}{
      \footnotemark[$*$] The peak bolometric flux of the normal bursts
      } \\
      \multicolumn{6}{l}{
      \footnotemark[$\dagger$] The ten-day average fluxes in the 2--20
      keV band.
      } \\
      \multicolumn{6}{l}{
      \footnotemark[$\ddagger$] 
       The references are 
       K10: \cite{2010A&A...514A..65K},
       G08: \cite{2008ApJS..179..360G},
       C02: \cite{2002A&A...392..885C}. 
       } \\
  \hline
\endlastfoot
     4U 0614+091    & no    
      & 4--33         & 103 $\pm$  6  & 135 $\pm$  4 & K10 \\
     SLX 1735$-$269 & yes   
      & 6             &  76 $\pm$  2  &  73 $\pm$  3 & G08 \\
     SAX J1747.0$-$2853 & yes 
      & 5.0$\pm$0.1   & 537 $\pm$  5  & 531 $\pm$  5 & G08 \\
     EXO 1745$-$248 & yes   
      & 6.0$\pm$0.1   &  55 $\pm$  2  & 130 $\pm$  3 & G08 \\
     4U 1820$-$30   & no    
      & 5.5$\pm$0.2   & 840 $\pm$  4  & 698 $\pm$ 10 & G08 \\
     SAX J1828.5$-$1037 & yes 
      & 4.3$\pm$1.6   &  18 $\pm$  2  &  30 $\pm$  2 & C02 \\
     4U 1850$-$086 (1st) & yes 
      & 5.9$\pm$0.9   &  12 $\pm$  3  &  22 $\pm$  4 & This work \\
     \phantom{4U 1850$-$086} (2nd)    &
      &               &   7 $\pm$  5  &  16 $\pm$  3  & {}  \\
     Aql X-1    & yes       
      & 8.9$\pm$1.5   & 956 $\pm$ 13  & 338 $\pm$  4  & G08 \\
\end{longtable}

\begin{longtable}{llllll}
  \caption{The spectral parameters of the bursts.}
  \label{tab:spec}
      \hline
      Name & MJD\footnotemark[$*$] 
        & $kT_\mathrm{bb}$\footnotemark[$\dagger$] 
          & $F_\mathrm{bb}$\footnotemark[$\ddagger$]  
           & $N_{\mathrm H}$\footnotemark[$\S$]   
            & $\chi^2$ (DoF)\footnotemark[$\|$] \\ 
        &     & (keV) & ($10^{-8}$ ergs cm$^{-2}$ s$^{-1}$)
      &  ($10^{22}$ cm$^{-2}$) & \\
      \hline
\endfirsthead
      \hline
      Name & MJD & $kT_\mathrm{bb}$ & $F_\mathrm{bb}$ 
      & $N_{\mathrm H}$ & $\chi^2$ (DoF)\\ 
        &     & (keV) & ($10^{-8}$ ergs cm$^{-2}$ s$^{-1}$)
      &  ($10^{22}$ cm$^{-2}$) & \\
      \hline
\endhead
  \hline
\endfoot
  \hline
    \multicolumn{6}{l}{
    \footnotemark[$*$] The center time of the scan in Modified Julian Date
    } \\
    \multicolumn{6}{l}{
    \footnotemark[$\dagger$] The blackbody temperature of the spectrum
    } \\
    \multicolumn{6}{l}{
    \footnotemark[$\ddagger$] The bolometric flux of the spectrum
    } \\
    \multicolumn{6}{l}{
    \footnotemark[$\S$] The hydrogen column density
    } \\
    \multicolumn{6}{l}{
    \footnotemark[$\|$] The $\chi^2$ and the degrees of freedom of the fit
    } \\
  \hline
\endlastfoot
      4U 0614+091        &&&& --- & 83.62 (84)  \\
       & 56964.63 & 1.9$_{-0.2}^{+0.2}$ & 3.3$_{-0.4}^{+0.4}$ 
        & & \\
       & 56964.69 & 1.5$_{-0.2}^{+0.2}$ & 1.6$_{-0.3}^{+0.3}$ 
        & &  \\
       & 56964.76 & 1.6$_{-0.3}^{+0.3}$ & 1.9$_{-0.4}^{+0.4}$ 
        & & \\
       & 56964.82 & 1.8$_{-0.5}^{+0.8}$ & 1.3$_{-0.4}^{+0.6}$ 
        & & \\
       & 56964.89 & 1.8$_{-0.5}^{+0.5}$ & 1.0$_{-0.3}^{+0.4}$ 
        & & \\
      SLX 1735$-$269    &&&& --- & 24.43 (29) \\
       & 56267.15 & 2.8$_{-0.3}^{+0.3}$ & 4.4$_{-0.6}^{+0.7}$ 
        & & \\
       & 56267.15 & 1.9$_{-0.5}^{+0.8}$ & 0.5$_{-0.2}^{+0.2}$ 
        & & \\
      SAX J1747.0$-$2853 &&&& 10 & 131.79 (143) \\
       & 55605.58 & 2.0$_{-0.3}^{+0.3}$ & 1.5$_{-0.2}^{+0.2}$
        & & \\
       & 55605.65 & 1.5$_{-0.3}^{+0.3}$ & 1.0$_{-0.2}^{+0.2}$
        & & \\
       & 55605.71 & 1.4$_{-0.4}^{+0.5}$ & 0.6$_{-0.2}^{+0.2}$
        & & \\
       & 55605.77--55605.84 & 1.2$_{-0.4}^{+0.4}$ & 0.5$_{-0.2}^{+0.2}$
        & & \\
      EXO 1745$-$248    &&&&&  \\
       & 55858.56 & 2.2$_{-0.3}^{+0.3}$ & 1.2$_{-0.2}^{+0.2}$
        & --- &  8.93 (16) \\
       & 55858.62 & 1.7$_{-0.4}^{+0.5}$ & 0.6$_{-0.2}^{+0.2}$
        & --- &  6.09 (8) \\
       & 55858.75 & 1.7$_{-0.2}^{+0.3}$ & 0.5$_{-0.1}^{+0.1}$
        & --- &  5.68 (6) \\
       & 55858.82--55858.88 & 1.6$_{-0.3}^{+0.5}$ & 0.3$_{-0.1}^{+0.1}$
        & --- &  3.59 (4) \\
       & 55858.94--55859.14 & 1.2$_{-0.3}^{+0.3}$ & 0.1$_{-0.03}^{+0.03}$
        & --- &  3.32 (4) \\
      4U 1820$-$30       &&&& --- & 223.29 (203) \\
       & 55272.72 & 3.0$_{-0.2}^{+0.2}$ & 5.3$_{-0.5}^{+0.5}$
        & &  \\
      SAX J1828.5$-$1037 &&&&&  \\
       & 55877.36 & 2.7$_{-0.6}^{+0.9}$ & 0.9$_{-0.3}^{+0.3}$
        & 4.1 &  3.03 (5) \\
       & 55877.43 & 1.3$_{-0.3}^{+0.3}$ & 0.4$_{-0.1}^{+0.1}$
        & 4.1 &  3.11 (5) \\
       & 55877.49--55877.55 & 1.5$_{-0.3}^{+0.3}$ & 0.2$_{-0.1}^{+0.1}$
        & 4.1 &  5.53 (6) \\
      4U 1850$-$086 (2nd) &&&&&  \\
       & 57151.39 & 2.6$_{-0.1}^{+0.1}$ & 7.8$_{-0.5}^{+0.5}$
        & --- & 90.49 (98) \\
       & 57151.46 & 0.7$_{-0.2}^{+0.3}$ & 0.2$_{-0.1}^{+0.1}$
        & --- & 0.82  (3) \\
      Aql X-1            &&&& --- & 173.15 (188) \\
       & 56493.30 & 1.8$_{-0.2}^{+0.2}$ & 2.0$_{-0.3}^{+0.3}$ 
        & & \\
       & 56493.37 & 1.9$_{-0.3}^{+0.3}$ & 1.2$_{-0.3}^{+0.3}$ 
        & & \\
       & 56493.43 & 1.6$_{-0.4}^{+0.6}$ & 0.5$_{-0.2}^{+0.2}$ 
        & & \\
       & 56493.50 & 1.4$_{-0.3}^{+0.4}$ & 0.6$_{-0.2}^{+0.2}$ 
        & & \\
       & 56493.56 & 1.2$_{-0.3}^{+0.4}$ & 0.5$_{-0.2}^{+0.2}$ 
        & & \\
       & 56493.63 & 1.9$_{-1.2}^{+1.6}$ & 0.5$_{-0.2}^{+0.3}$ 
        & & \\
       & 56493.69 & 2.3$_{-1.1}^{*}$  & 0.4$_{-0.2}^{+6.3}$ 
        & & \\
       & 56493.76 & 1.0$_{-0.4}^{+0.5}$ & 0.3$_{-0.2}^{+0.2}$ 
        & & \\
\end{longtable}

\begin{longtable}{*{7}{l}}
  \caption{The properties of the bursts studied in this paper.}
    \label{tab:bst}
      \hline
      Name & 
      $F_\mathrm{obs}$ & $F_\mathrm{max}$ & $\tau_\mathrm{LB}$ & 
      $d$ & $E_{\mathrm b}$\footnotemark[$*$] 
      & first burst\footnotemark[$\dagger$] \\ 
      & \multicolumn{2}{l}{($10^{-8}$ ergs cm$^{-2}$ s$^{-1}$)} & 
      (hour) & (kpc) & ($10^{41}$ ergs) & (day)\\
      \hline
\endfirsthead
      \hline
      Name & 
      $F_\mathrm{obs}$ & $F_\mathrm{max}$ & $\tau_\mathrm{LB}$ & 
      $d$ & $E_{\mathrm b}$\footnotemark[$*$] 
      & first burst\footnotemark[$\dagger$] \\
      & \multicolumn{2}{l}{($10^{-8}$ ergs cm$^{-2}$ s$^{-1}$)} & 
      (hour) & (kpc) & ($10^{41}$ ergs) & (day)\\
      \hline
\endhead
  \hline
\endfoot
  \hline
  \multicolumn{7}{l}{
    \footnotemark[$*$] 
    The radiated burst energy with statistic (systematic) error.
    } \\
  \multicolumn{7}{l}{
    The systematic error is due to the uncertainty of the peak flux.
    } \\
  \multicolumn{7}{l}{
    \footnotemark[$\dagger$] 
    The time of the first X-ray burst after the sample event.
    } \\
  \multicolumn{7}{l}{
    \footnotemark[$\ddagger$] 
    The number in the parentheses is the 
    observed peak flux by INTEGRAL with bolometric correction (see text).
    } \\
  \multicolumn{7}{l}{
    References are 
    (A) \cite{1992A&A...262L..15B}, (B) \cite{2010A&A...514A..65K},
    (C) \cite{2008ApJS..179..360G}, 
    } \\
  \multicolumn{7}{l}{
    (D) \citet{2011ATel.3183....1C},
    (E) \cite{2000ApJ...543L..73N},
    (F) \cite{2011ATel.3217....1L},
    (G) \cite{2007A&A...470.1043O}, 
    } \\
  \multicolumn{7}{l}{
    (H) \cite{2010arXiv1012.3224H}, 
    (I) \cite{2011ATel.3625....1I},
    (J) \cite{2002A&A...392..885C},
    (K) \cite{2014ATel.5972....1I}.
    } \\
  \hline
\endlastfoot
      4U 0614+091        
        & 3.3$_{-0.4}^{+0.4}$ & 4.0 & 5.2 & 3 (A,B)  
         & 6.7 $\pm$0.8 (+1.4) & --- \\
      SLX 1735$-$269     
        & 4.4$_{-0.6}^{+0.7}$ & 32 & 0.77 & 7.3 (C)   
         & 7.8 $\pm$1.2 (+49) & --- \\
      SAX J1747.0$-$2853 
        & 1.5$_{-0.2}^{+0.2}$ & 1.9 (6.9)\footnotemark[$\ddagger$](D)  
        & 4.2 & 9 (E)     & 22 $\pm$2.9 (+5.4) & $\leq$25 (F)\\
      EXO 1745$-$248    &  1.2$_{-0.2}^{+0.2}$ 
        & 1.5 & 4.2 & 5.5 (G) & 6.6 $\pm$1.1 (+1.7) & --- \\
      & & &         & 6.9 (H) &  10 $\pm$1.7 (+2.6) &  \\
      4U 1820$-$30       & 5.3$_{-0.5}^{+0.5}$ 
        & 110 & 0.5$\pm$0.1 (I) & 7.9 (H) 
         & 7.1 $\pm$2.1 (+146)  & $\leq$1549 (this work) \\
      SAX J1828.5$-$1037 & 0.9$_{-0.3}^{+0.3}$ 
        & 1.7 & 2.3 & $<$6.2 (J) & $<$3.4 $\pm$1.1 (+3.2) & --- \\
      4U 1850$-$086 (1st)  
        & 10.7$\pm$0.9 (K) & --- & 0.27 (K) & 6.9 (H) 
         & 5.9 $\pm$0.5 (+0) & $\leq$425 (this work) \\
      \hspace{13ex}  (2nd) 
        & 7.8$_{-0.5}^{+0.5}$ & 13  &  0.71   &       
         & 11 $\pm$0.7 (+7.8)  & $\leq$186 (this work) \\
      Aql X-1            
        & 2.0$_{-0.3}^{+0.3}$ & 2.6 & 4.3 & 5.0 (C)  
         & 9.3 $\pm$1.4 (+2.7) & $\leq$389 (this work) \\
\end{longtable}


\begin{ack}
This research has made use of the MAXI data provided by RIKEN, JAXA 
and the MAXI team.
This research was supported by JSPS KAKENHI
Grant Number 24740186, JP16K17717.
\end{ack}


\bibliographystyle{aa}
\bibliography{ref}

\end{document}